\documentclass[npb,preprint,showpacs,preprintnumbers,nofootinbib,amsmath]{revtex4}
\usepackage{graphicx}

\newcommand{\bit}[1]{\mbox{\boldmath$#1$}}

\newcommand{\vecc}[1]{\mbox{\boldmath $#1$}}

\def\({\left(}
\def\[{\left[}
\def\){\right)}
\def\]{\right]}
\def\pd{\partial}

\begin{document}
\preprint{\hbox{RUB-TPII-01/2010}}

\title{Wilson lines in transverse-momentum dependent parton distribution
       functions with spin degrees of freedom}
\author{I.~O.~Cherednikov}
\email{igor.cherednikov@jinr.ru}
\affiliation{INFN Cosenza, Universit$\grave{a}$
             della Calabria, I-87036 Arcavacata di Rende (CS), Italy\\
             and \\
             Bogoliubov Laboratory of Theoretical Physics,
             JINR,
             RU-141980 Dubna, Russia\\}
\affiliation{ITPM,
             Moscow State University, RU-119899 Moscow, Russia\\}
\author{A.~I.~Karanikas}
\email{akaran@phys.uoa.gr}
\affiliation{University of Athens, Department of Physics,
             Nuclear and Particle Physics Section,
             Panepistimiopolis, GR-15771 Athens, Greece\\}
\author{N.~G.~Stefanis}
\email{stefanis@tp2.ruhr-uni-bochum.de}
\affiliation{Institut f\"{u}r Theoretische Physik II,
             Ruhr-Universit\"{a}t Bochum,
             D-44780 Bochum, Germany\\}
\affiliation{Bogoliubov Laboratory of Theoretical Physics,
             JINR,
             RU-141980 Dubna, Russia}

\date{\today}

\begin{abstract}
We propose a new framework for transverse-momentum dependent parton
distribution functions, based on a generalized conception of gauge
invariance which includes into the Wilson lines the Pauli term
$\sim F^{\mu\nu}[\gamma_\mu, \gamma_\nu]$.
We discuss the relevance of this nonminimal term for unintegrated
parton distribution functions, pertaining to spinning particles,
and analyze its influence on their renormalization-group properties.
It is shown that while the Pauli term preserves the probabilistic
interpretation of twist-two distributions---unpolarized and
polarized---it gives rise to additional pole contributions to those
of twist-three.
The anomalous dimension induced this way is a matrix, calling for a
careful analysis of evolution effects.
Moreover, it turns out that the crosstalk between the Pauli term and
the longitudinal and the transverse parts of the gauge fields,
accompanying the fermions, induces a constant, but process-dependent,
phase which is the same for leading and subleading distribution
functions.
We include Feynman rules for the calculation with gauge links
containing the Pauli term and comment on the phenomenological
implications of our approach.
\end{abstract}


\maketitle

\section{Introduction}
\label{sec:intro}

Parton distribution functions (PDF)s are the key nonperturbative
ingredients of completely inclusive QCD processes, like deeply
inelastic scattering (DIS).
The process-dependent hard-scattering part of such processes
can be calculated order by order in QCD perturbation theory on
account of the hard scale of the process
$Q^2\gg \Lambda_{\rm QCD}^2$.
Though the determination of the initial PDF requires the application
of nonperturbative methods, its $Q^2$ evolution is controlled by
renormalization-group (RG) evolution equations with anomalous
dimensions calculable within perturbative QCD.

This simple picture changes significantly when one considers
semi-inclusive processes, like semi-inclusive DIS (SIDIS),
or the Drell-Yan (DY) process in hadronic collisions, in which
hadrons are detected in the final (initial) state with a sizeable
transverse momentum.
In that case, one needs information about the generation of the
transverse momentum $P^h_\perp$ of the final (initial) hadrons, e.g.,
by means of the transverse-momentum distribution of the partons.
This mechanism is believed to be dominant at small $P^h_\perp \ll Q$,
while at large $P^h_\perp \sim Q$, the transverse momentum $P^h_\perp$
is produced by the perturbative gluon exchanges.
The second mechanism, as well as the relationship between the two in
the intermediate region, are outside the scope of the present work.
In any case, integrated PDFs of leading twist are not sufficient to
describe semi-inclusive processes.
One therefore introduces transverse-momentum dependent (TMD) PDFs
which keep track of the intrinsic transverse motion of the partons
inside the hadrons and reveal this way fine details about their
substructure
(pioneering works are \cite{S76,S79,CS81,CS82}---see also
\cite{Col03,TRENTO04,BR05,DM07,Bacch08,STM10} and Refs.\ cited
therein, and \cite{CSS89} for a review).
The introduction of TMD PDFs, though intuitively clear and physically
appealing, still poses serious challenges.
The first problem is related to the TMD factorization: Its status
beyond leading twist (and to all orders) is far from being satisfactory
at the moment \cite{JMY04,CRS07,CQ07,RM2010}.
Next, there is a possible \textit{non-universality} of TMD PDFs
entailed by the extremely complicated and often process-dependent
structure of the gauge links,\footnote{These are path-ordered
exponentials of the gauge field, needed to render the definition of
PDFs gauge invariant.} see Refs.\ \cite{CM04,BMP03,BM07}.
Finally, in the light-cone gauge, \textit{extra divergences} appear
that have to be properly treated
\cite{CS82,Col03,Col08,CS09-ER}---in contrast to the integrated
case.
In the present paper, we focus on the last two issues.

In the integrated case, the parton density $f_{i/h}(x,Q^2)$
describes the probability to find a parton $i$ with longitudinal
momentum fraction $xP^+$ inside hadron $h$ with momentum $P$,
and can be given a gauge-invariant definition in terms of the gauge
link (Wilson line) [see, for instance, \cite{ER80Riv}]
\begin{equation}
  [\xi^-;0^-|\Gamma]
=
  \mathcal{P} \exp
                  \left[
                  - ig \int_{0^{-}[\mathcal{C}]}^{\xi^-}
                           dz^\mu A_{\mu}^{a}(z)t^a
                  \right]
\label{eq:gaugelink}
\end{equation}
for a contour $\mathcal{C}$ along the light-cone, where the
path-ordered exponential $A_{\mu}^{a}$ refers to the (gluon)
gauge field.
The renormalization of the integrated PDF obeys the
Dokshitzer-Gribov-Lipatov-Altarelli-Parisi (DGLAP)
\cite{DGLAP,AP77} evolution equation with its integral kernel being
related to the anomalous dimension
$
 \Gamma_{i/h}
=
 \Sigma_{\rm q}\Gamma_{q} + 2\, \Gamma_{\rm end}
$,
where $\Gamma_{\rm end}$ is the endpoint anomalous dimension
of the integration contour $\mathcal{C}$ in the gauge link (see
\cite{CS08} for a more detailed discussion of this issue and
\cite{CD80,Aoy81,Ste83} for the original derivations
and earlier references).

It was pointed out in \cite{JY02,BJY03,BMP03} that a completely
gauge-invariant definition of the TMD PDF in those gauges
in which $\mathbf{A}_\perp$ does not vanish at infinity has to include
also transverse gauge links.
Hence, in the light-cone gauge $A^+=0$, applied in conjunction with
$q^-$-independent pole prescriptions (like the advanced, retarded or
principal-value prescription) in order to avoid singularities at $q^+$
in the gluon propagator, the transverse gauge links receive radiative
corrections that can be associated with a cusp-like junction point at
light-cone infinity \cite{CS07,CS08,SC09}.
The emerging cusp anomalous dimension \cite{KR87} has to be removed,
if one aims to recover the results valid in covariant gauges.
To this end, a redefinition of the TMD PDF was proposed by two of us
\cite{CS07,CS08} which involves a soft factor, termed $R$, consisting
of two eikonal lines evaluated along a particular gauge (integration)
contour with a jackknifed path segment in the transverse direction
(see next section).
Note that the introduction of the soft factor can be justified from a
different point of view as well---see, for instance, Refs.\
\cite{CH00,Hau07,CM04}, where the soft factor was used to take care of
rapidity divergences in covariant gauges.
The anomalous dimension related to the ultraviolet (UV) divergences
(e.g., pole terms in $1/\epsilon$ in dimensional regularization)
of the soft factor was found \cite{CS07,CS08} to exactly compensate
(at the one-loop order) the cusp anomalous dimension of the transverse
gauge link, hence, ensuring the independence of the (redefined) TMD
PDF from artificial contour-generated anomalous-dimension artifacts.

More recently \cite{CS09}, we have shown that this factorization
scheme remains valid also for the Mandelstam-Leibbrandt pole
prescription \cite{Man82,Lei83}, which is $q^-$-dependent.
In that case, the UV divergent part of the soft factor reduces to
unity, while the transverse gauge link does not give rise to a defect
of the anomalous-dimension that has to be compensated.
As a result, the TMD PDF has the same anomalous dimension as in
covariant gauges, rendering the proposed definition of the TMD PDF
gauge and pole-prescription independent.

\begin{figure*}[ht]
\centering
\includegraphics[width=0.30\textwidth,angle=90]{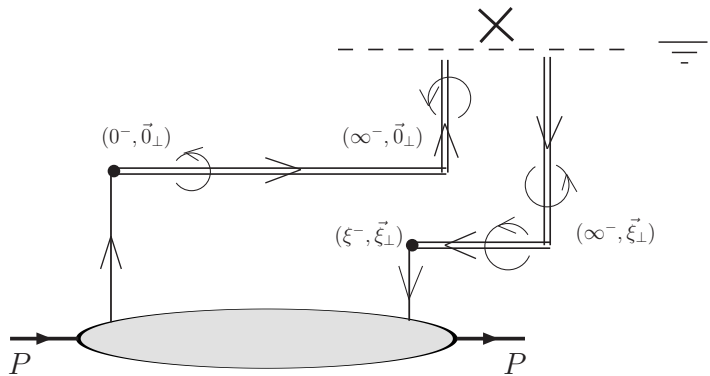}~~
\caption{Graphical representation of a generic TMD PDF
in coordinate space with the spin-dependent Pauli terms
included in the gauge links.
The (hidden) contour obstruction at transverse and lightlike infinity
(``earth'' symbol) is represented by $\times$.
\protect\label{fig:TMD_pauli}}
\end{figure*}

The basic tenet in the gauge-invariant formulation of hadronic
quantities, like TMD PDFs, is to use a gauge link with an exponent
which contains only the gauge field $A$.
However, this is only the simplest (or minimal-twist) possibility
which pays attention to the fact that color vectors
cannot be compared at a distance.
Because the gauge potential $A_{a}^{\mu}$ is spin-blind, one should
actually include into the gauge link an additional term proportional
to the gluon tensor $F_{\mu\nu}^{a}$---called the Pauli term---which
can accommodate the direct spin-dependent interaction in accordance
to the Lorentz group.
This term represents the \textit{minimal} coupling of a spinning
particle to an external field and may become important for
non-trivial contours, while additional terms of still higher
twist are not prohibited but are relatively power-suppressed.
Thus, the gauge links will be generalized to take into account
the Pauli contribution \cite{Ste-unpub}
$\sim F_{\mu\nu}^{a} S_{\mu\nu}$, where
$S_{\mu\nu}=(1/4)[\gamma_\mu, \gamma_\nu]$,
normally ignored.
This means that in order to accommodate spin-dependent interactions
in a manifestly gauge-invariant formalism, one has to include the
following path-ordered exponential
$
 \mathcal{P}\exp\left[-i g \int_{0}^{\infty}
 d\sigma S_{\mu\nu} F^{\mu\nu}_{a}(u\sigma)t^{a} \right]
$.
The graphic illustration of this concept is depicted in simple
contextual terms in Fig.\ \ref{fig:TMD_pauli} which shows a generic
process with gauge links that contain the Pauli term---codified by
small rings around the double lines which stand for the conventional
gauge links.

It is expected that any effects of such spin-dependent terms should be
non-vanishing only in the case of (at least) transverse gauge links off
the light-cone.
In the integrated collinear PDFs, the Wilson lines are one-dimensional
in the sense that the paths of the integration reduce to lightlike
rays.
Going beyond the fully collinear picture, which is unavoidable in
semi-inclusive processes, one must make use of gauge links which
involve more complicated integration contours, which have at least one
additional---transverse---dimension.
Note that in the case of integration paths off the light-cone---used
to regularize rapidity divergences \cite{CS82}---one has, in fact, even
more non-trivial contours because they contain, beyond the minus
light-cone segments, also plus components.
Therefore, in the TMD case, effects related to the spin transfer
from the starting point, say, $(0^-, \vecc 0_\perp)$, to the
terminating point $(\xi^-, \vecc \xi_\perp)$ may become (at least, in
principle) apparent due to the non-trivial structure of the contour.
Hence the cross-talk between a pair of quantum fields at distant points
off the light-cone will contain spin-dependent Pauli terms which are of
higher twist order with respect to the spin-blind ones containing the
gauge potential.
Our analysis reveals that the inclusion of the Pauli term, although
non-visible in the completely unpolarized TMD PDFs, can produce
non-vanishing effects in a number of polarized distributions,
in particular, those responsible for time-reversal-odd
phenomena.\footnote{We thank A.~V.\ Efremov for important comments
on this point.}

Adopting this encompassing concept of gauge invariance, questions
arise whether the definition of TMD PDFs, we proposed before
in Refs.\ \cite{CS07,CS08}, has to be modified and whether the
inclusion of the Pauli term has phenomenological consequences---as
already indicated.
The first issue is related to the question whether spin-dependent
terms affect the factorization schemes discussed in our previous works,
while the second addresses possible changes of the RG, i.e., evolution
behavior of TMD PDFs.
The present work is devoted to the clarification of these issues.

The rest of the paper is organized as follows.
The pivotal Sec.\ \ref{sec:spin-effects} argues that the correct
treatment of spin degrees of freedom in the TMD PDFs necessitates the
inclusion into the gauge links of the Pauli term.
This contribution describes the interaction between spinning particles
and the gauge-field strength and leads to a generalization of Eq.\
(\ref{eq:gaugelink}).
Its implications are worked out in Sec.\ \ref{sec:pauli}.
The calculation of virtual gauge-field correlators for the
leading-twist distributions, as well as for those of subleading twist,
is carried out in Sec.\ \ref{sec:virtual-corr}, whereas those related
to fermions are discussed in Sec.\ \ref{sec:fermion-virt}.
Section \ref{sec:real-contr} is concerned with the consideration of
contributions to the TMD PDFs stemming from real-gluon emission.
Finally, in Sec.\ \ref{sec:concl}, we summarize the results and present
our conclusions.
To go further with QCD calculations with gauge links, which include the
Pauli term, we develop a set of Feynman rules and display them in Fig.\
\ref{fig:pauli-feynman}.

\section{Inclusion of spin effects}
\label{sec:spin-effects}

The TMD PDF for an unpolarized/polarized quark of flavor $i$ in an
unpolarized/polarized target $h$ following our generalized concept of
gauge invariance reads
\begin{eqnarray}
\begin{split}
  f_{i/h}^{\Gamma}(x, \mbox{\boldmath$k_\perp$})
= {}&
   \frac{1}{2} {\rm Tr} \! \int\! dk^-
   \int \! \frac{d^4 \xi }{(2\pi)^4}\,
   {\rm e}^{- i k \cdot \xi}  \
   \langle  h \ |\bar \psi_i (\xi) \\
&  \times [[\xi^-, \mbox{\boldmath$\xi_\perp$};
   \infty^-, \mbox{\boldmath$\xi_\perp$}]]^\dagger
   [[\infty^-, \mbox{\boldmath$\xi_\perp$};
   \infty^-, \mbox{\boldmath$\infty_\perp$}]]^\dagger \cdot \Gamma  \\
&  \left.
   \times [[\infty^-, \mbox{\boldmath$\infty_\perp$};
   \infty^-, \mbox{\boldmath$0_\perp$}]]
   [[\infty^-, \mbox{\boldmath$0_\perp$};
   0^-, \mbox{\boldmath$0_\perp$}]]
\psi_i (0) | h \right \rangle \! \cdot \!
   R
\vspace{-1cm}
\label{eq:TMD-PDF}
\end{split}
\end{eqnarray}
where $\Gamma$ denotes one or more $\gamma$ matrices in correspondence
with the particular distribution in question, and the state
$|h\rangle$ stands for the appropriate target.
In the unpolarized case, we have $|h\rangle=|h(P)\rangle$, with $P$
being the momentum of the initial hadron, whereas for a (transversely)
polarized target the state is
$|h\rangle=|h(P), S_{\perp}\rangle$.
The ``enhanced'' gauge links $[[\xi_2;\xi_1]]$ and the soft factor
$R$ will be defined shortly.

An important comment about definition (\ref{eq:TMD-PDF}) is here in
order before we proceed.
We started from the ``fully unintegrated'' correlation function, which
depends on \textit{all} four components of the parton's momentum
\cite{TM94,CRS07}.
Thus, the TMD PDF is obtained \textit{after} performing the $k^-$
integration, which formally renders the coordinate $\xi^+$ equal to
zero:
\begin{equation}
  \int\! dk^- {\rm e}^{- i k^- \xi^+}
=
  2\pi \delta (\xi^+) \ .
\label{eq:delta_reg}
\end{equation}
However, one must be careful:
This operation may produce additional divergences because, carrying it
out, all quantum fields involved (quarks and gluons) will be defined
on the light ray $\xi^+ = 0$.
This means that the plus light-cone coordinates of the product
of two quantum fields will always coincide.
To avoid this, we will regularize this singularity in what follows by
taking into account that a particle, once created at the point
$\xi^+=0$, will be reabsorbed (destroyed) with the \textit{same}
probability at (potentially very distant) points
$0\pm \Delta/2$, where
$\Delta \sim 1/p^- \sim p^+/2M^2$
is the uncertainty of determining a point along the plus direction.
In other words, we have to sum (average) over all indistinguishable
possibilities in order to get the correct answer in the quantum
mechanical sense.
For instance, the regularized two-gluon correlator is written as
\begin{equation}
\begin{split}
& \frac{1}{T} \int_{-\Delta /2}^{\Delta /2}\! dt \
  \left\langle {\cal A}^\mu (0^+, \xi^-, \vecc \xi_\perp)
               {\cal A}^\nu (t, \xi'^-, \vecc \xi '_\perp)
  \right\rangle{_{_0}}
\stackrel{\Delta \to \infty}{=}
  \frac{-i C_{\rm F}}{T}
  \int\!  \frac{d^4 q}{\( 2\pi \)^4 } \
  {\rm e}^{
           -i q^+\(\xi^- - \xi '^- \) + i \bit{\scriptstyle q_\perp}
           \cdot \( \bit{\scriptstyle \xi_\perp}
           - \bit{\scriptstyle \xi '_\perp}
                 \)
           } \\
& ~~~~~~~~~~~~~~~~~~~~~~~~~~ ~~~~~~~~~~~~~~~~~~~~~~~~~~ ~~~~~~~~~~~ \times
  2\pi \delta \(q^-\) D^{\mu\nu}(q) \ ,
\label{eq:regular_pauli}
\end{split}
\end{equation}
whereas without regularization, the corresponding term
$\sim \int\! dq^- \ D^{\mu\nu} (q^+,q^-,\vecc q_\perp)$ would
face unphysical UV divergences.

The constant $T \sim 1/p^+$ (so to say the ``length'' of the plus ray)
will drop out from all final results, provided a suitable
parametrization of the vectors along the contour integral is adopted.
This is crucial for the enhanced gauge link which includes the Pauli
term, since the latter is not reparameterization invariant---in
contrast to the usual gauge link.
Therefore, we make use of the following reparameterization of the
(initially dimensionless) constant vectors that define the motion along
the line integral:
\begin{equation}
  n^+_\mu \to u^*_\mu
=
  p^- n^+_\mu \ , \quad
  n^-_\mu
\rightarrow
  u_\mu
=
  p^+ n^-_\mu \ , \quad
  \vecc l_{\perp}
\rightarrow
  p^+ \vecc l_{\perp} \ ,
\label{eq:rescaling}
\end{equation}
which implies boosts in the collinear directions.
Note that the plus-component of the momentum, $p^+$, is large in our
kinematics and is the only mass scale entering the above
reparameterization.
Thus, the uncertainty of determining a position along the plus ray in
Eq.\ (\ref{eq:regular_pauli}) is very large, namely,
$$\Delta \sim \frac{1}{u^*} = \frac{1}{p^-}\ , $$
while it is very small along the minus or the transverse directions:
$$
 \Lambda
\sim
 \frac{1}{u}
\sim
 \frac{1}{|\vecc l_{\perp}|}
\sim
 \frac{1}{p^+} \ .
$$

We can now define the enhanced lightlike gauge link along the
$x^-$ direction:
\begin{eqnarray}
  [[\infty^-, \vecc 0_{\perp}; 0^-, \vecc 0_\perp]]
& = &
  \mathcal{P}
  \exp
      \left[
            - ig \int_{0}^{\infty} d\sigma \ u_{\mu} \
                 A_{a}^{\mu}(u \sigma)t^a
            - i g \int_{0}^{\infty} d\sigma  \
                 S_{\mu\nu} F_{a}^{\mu\nu}(u \sigma)t^a
      \right] \ .
\label{eq:lightlike-link}
\end{eqnarray}
An analogous definition holds for the $x^+$ direction by making the
replacement $u \to u^*$.
On the other hand, the enhanced transverse gauge link is given by
\begin{eqnarray}
  [[\infty^-,  \vecc\infty_{\perp}; \infty^-, \vecc 0_\perp]]
& = &
  \mathcal{P}
  \exp
      \left[
            - ig \int_{0}^{\infty} d\tau
            \vecc l_{\perp} \! \cdot \!
            \vecc A_{\perp}^{a}(\vecc l\tau)t^a
            - i g \int_{0}^{\infty} d\tau
            S_{\mu\nu}F_{a}^{\mu\nu}(\vecc l\tau)t^a
      \right] \ ,
\label{eq:transverse-link}
\end{eqnarray}
where the two-dimensional vector
$\mbox{\boldmath $l$}\equiv \vecc l_\perp$
drops out from all final results, and the Lorentz generators for
the spin are defined by
$S_{\mu\nu}=(1/4)[\gamma_{\mu},\gamma_{\nu}]$.
Note that the path ordering, denoted by $\mathcal{P}$ in the
compound expressions above, means
\begin{widetext}
\begin{equation}
\begin{split}
&  \mathcal{P}[\ldots]
=
  1 + \int_{0}^{\infty} d\tau_{1}
  \mathcal{P} \exp \left(
                         \int_{\tau_{1}}^{\infty} d\tau
                         u \cdot \mathcal{A}
                   \right)
  g S \cdot \mathcal{F}(u \tau_{1})
  \mathcal{P} \exp \left(
                          \int_{0}^{\tau_{1}} d\tau
                          u \cdot \mathcal{A}
                   \right) \\
& ~~~~~~~~~~~~~ +
  \int_{0}^{\infty}\! d\tau_{2}
     \int_{0}^{\tau_{2}} d\tau_{1}
  \mathcal{P} \exp \left(
                             \int_{\tau_{2}}^{\infty} d\tau
                             u \cdot \mathcal{A}
                       \right) \cdot
    g S \cdot \mathcal{F}( u \tau_{2})
    \mathcal{P} \exp \left(
                            \int_{\tau_{1}}^{\tau_{2}} d\tau
                            u \cdot \mathcal{A}
                     \right) \\
& ~~~~~~~~~~~~~ \times
    g S \cdot \mathcal{F}(u \tau_{1})\cdot
    \mathcal{P} \exp \left(
                            \int_{0}^{\tau_{1}} d\tau
                            u \cdot \mathcal{A}
                     \right)
    + \ldots \ ,
\label{eq:path-ordering}
\end{split}
\end{equation}
\end{widetext}
where we have used the following convenient abbreviations:
$\mathcal{A}=\sum_{a}A_{a}t^{a}$,
$u \cdot A = \sum_{\mu} u_{\mu}A^{\mu}$,
$S \cdot F= \sum_{\mu,\nu}S_{\mu\nu} F^{\mu\nu}$,
and the path ordering inside Eq.\ (\ref{eq:path-ordering}) is the
usual one.
It becomes obvious that the enhanced gauge links, defined above, and
the standard ones fulfil the same gauge transformations.

\begin{figure*}[t]
\centering
\includegraphics[scale=0.55,angle=90]{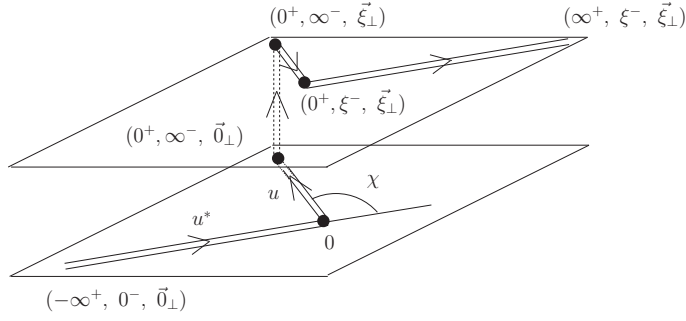}~~
\caption{Integration contour associated with the soft factor $R$ in
Eq.\ (\protect\ref{eq:R-factor}).
The cusp angle $\chi$ is explicitly indicated.
\protect\label{fig:R-contour}}
\end{figure*}

The soft factor $R$ in Eq.\ (\ref{eq:path-ordering})---introduced in
\cite{CS07,CS08} with the aim to remove the defect of the anomalous
dimension of the TMD PDF---may, in principle, be upgraded to include
the tensor term as well.
This amounts to the following expression
\begin{widetext}
\begin{equation}
\begin{split}
&   R(p^{+}, n^{-}|\xi^{-}, \vecc \xi_{\perp})
=
  {\rm Tr} \ \left\langle 0 \left|
     \mathcal{P} \exp
                      \left[
                             ig \int_{C_{\rm cusp}} ds\
                             \dot{\zeta} \cdot \mathcal{A}(\zeta)
                            +i g \int_{C_{\rm cusp}} ds\
                             S \cdot \mathcal{F}(\zeta)
                      \right] \right. \right. \\
& \left.\left.  ~~~~~~~~~~~~~ ~~~~~~~~~~~ \times
     \mathcal{\bar P} \exp
                      \left[
                            -ig \int_{C_{\rm cusp}} ds\
                             \dot{\zeta} \cdot \mathcal{A}(\xi+\zeta)
                            - i g \int_{C_{\rm cusp}} ds\
                             S \cdot \mathcal{F}(\xi+\zeta)
                      \right]
    \right| 0 \right\rangle \ ,
\label{eq:R-factor}
\end{split}
\end{equation}
\end{widetext}
where $\dot{\zeta}(s)=d\zeta/ds$, $\mathcal{\bar P}$ denotes anti-path
ordering, and the integration contour $C_{\rm cusp}$ is the same as
that employed in \cite{CS07,CS08}
(see Fig.\ \ref{fig:R-contour} for an illustration).
Note in this context that the soft factor was introduced before
(without the Pauli term) in Refs.\ \cite{CH00,Hau07} with the purpose to
control rapidity divergences of non-lightlike Wilson lines in
covariant gauges.
The soft factors in both approaches are multiplicative renormalization
eikonal factors, though in \cite{CH00,Hau07} the contribution from the
gauge link at infinity is not considered owing to the use of a
covariant gauge.

\section{Influence of the Pauli term}
\label{sec:pauli}

Before we focus our attention to the specific implications of the Pauli
term, let us first summarize the key features of the proposed scheme.
The usefulness of Eq.\ (\ref{eq:TMD-PDF}) derives from the fact that
by virtue of the soft factor $R$ all gauge-dependent
anomalous-dimension artifacts, potentially contributing to the TMD PDF,
are absent \textit{ab initio} \cite{CS07,CS08} so that, integrating
over the transverse momenta, one obtains a PDF which is controlled by
the DGLAP evolution equation \cite{CS08,CS09-ER} with the usual
anomalous dimension.
Moreover, to this definition all pole prescriptions adopted to
evaluate the gluon propagator in the light-cone gauge are fungible
\cite{CS09}.

To study the effects of the spin-dependent terms, induced by the
inclusion of the Pauli contribution, it suffices to take them into
account only in the fermionic part of Eq.\ (\ref{eq:TMD-PDF}), leaving
the soft factor unmodified.
The justification of this treatment is based on the fact that the
structure of the soft factor is practically prescribed by the RG
properties of the {\it unsubtracted} TMD PDF, as shown in detail in
Refs.\ \cite{CS07,CS08,CS09}.
To be more specific, it was found there that the particular contour
$C_{\rm cusp}$ in the soft factor, depicted in Fig.\
\ref{fig:R-contour}, pertains to the cusp-like UV singularities
of the fermionic part of Eq.\ (\ref{eq:TMD-PDF}).
Another argument of retaining the original form of the soft factor
unchanged is provided by the requirement that it should be boost
invariant (see, e.g., Ref. \cite{Bacch08}).
Given that the Pauli term is not invariant under scale transformations,
we refrain from including it into the soft factor in the present
investigation.
From the calculational point of view, the above argument is related to
the fact that, in the absence of any Lorentz structure, the spin-field
interaction cannot produce nontrivial results for integration paths
without self-intersections---this will be considered elsewhere.

Carrying out the $k^-$ integration in Eq.\ (\ref{eq:TMD-PDF})
and leaving out the soft factor $R$, one obtains the following
\textit{unsubtracted} TMD PDF
\begin{widetext}
\begin{equation}
\begin{split}
   f_{i/q}^{\Gamma}(x, \mbox{\boldmath$k_\perp$})
 ={} &
   \frac{1}{2} {\rm Tr} \!
   \int \! \frac{d\xi^- d^2
   \mbox{\boldmath$\xi_\perp$}}{2\pi (2\pi)^2}\,
   \exp\left(- i k^+ \xi^- \!
   +\!  i \mbox{\boldmath$k_\perp$}
\cdot \mbox{\boldmath$\xi_\perp$}\right)
   \left\langle  p , s |\bar \psi_i (\xi^-, \mathbf{\xi}_\perp)
   [[\xi^-, \mbox{\boldmath$\xi_\perp$};
   \infty^-, \mbox{\boldmath$\xi_\perp$}]]^\dagger \right.\\
& \times [[\infty^-, \mbox{\boldmath$\xi_\perp$};
   \left. \infty^-, \mbox{\boldmath$\infty_\perp$}]]^\dagger
   \Gamma
   [[\infty^-, \mbox{\boldmath$\infty_\perp$};
   \infty^-, \mbox{\boldmath$0_\perp$}]]
   [[\infty^-, \mbox{\boldmath$0_\perp$};
   0^-, \mbox{\boldmath$0_\perp$}]] \right. \\
& \left. \times \,
   \psi_i (0^-,\mbox{\boldmath$0_\perp$}) | p, s \right \rangle \ .
\vspace{-1cm}
\label{eq:TMD-PDF_fermion}
\end{split}
\end{equation}
\end{widetext}

For our concrete calculations to follow, we consider matrix elements
between quark states having momentum $p$ and spin $s$:
$| p, s \rangle$.
Using the fermionic density matrix ($\hat p\equiv p \cdot \gamma$)
\begin{equation}
  u(k) \otimes \bar u (k)
  =
  \frac{1}{2} (\hat k + m) \(1 + \gamma_5 \hat s \) \ , \ s^2 = - 1 \ ,
\label{eq:density}
\end{equation}
with the spin vector
$s^\mu = (s^+,s^-,\vecc s_\perp)$
being given by \cite{TM94}
\begin{equation}
  s^\mu
=
  \lambda \( \frac{k^+}{m}, \frac{\vecc k_\perp^2 - m^2}{2 m k^+},
             \frac{\vecc k_\perp}{m} \)
  +
  \(0^+, \frac{\vecc k_\perp \cdot\vecc s_\perp}{k^+}, \vecc s_\perp \)
   \ ,
\end{equation}
one obtains in the tree-approximation
(indicated by the subscript 0 for $\alpha_{s}^{0}$)
\begin{equation}
\begin{split}
&
  f_{(0)}^{\Gamma}(x, \mbox{\boldmath$k_\perp$})
=
  \frac{1}{2}
  {\rm Tr} \[ (\hat p + m) \(1 + \gamma_5 {\hat s} \) \ \Gamma \] \
  \delta(p^+ - xp^+) \delta^{(2)} (\vecc k_\perp)
  \ .
\end{split}
\end{equation}
In particular, for the unpolarized TMD PDF with $\Gamma = \gamma^+$,
one has at leading twist two the expression
\begin{equation}
\begin{split}
&
  f_1^{(0)}
\equiv
  f_{(0)}^{\gamma^+}(x, \mbox{\boldmath$k_\perp$})
=
  \frac{1}{2}
  {\rm Tr} \[ (\hat p + m) \(1 + \gamma_5 {\hat s} \) \gamma^+ \]
  \delta(p^+ - xp^+) \delta^{(2)} (\vecc k_\perp) \\
&
 ~~~~~ =
  \delta(1-x) \delta^{(2)} (\vecc k_\perp)
  \ .
\end{split}
\end{equation}
On the other hand, the helicity and the transversity distributions are
given, respectively, by
\begin{equation}
\begin{split}
&
  f_{(0)}^{\gamma^+\gamma_5}(x, \mbox{\boldmath$k_\perp$})
=
  \delta(1-x) \delta^{(2)} (\vecc k_\perp) \ \lambda
  \ , \\
&
  f_{(0)}^{i \sigma^{i+}\gamma_5}(x, \mbox{\boldmath$k_\perp$})
=
  \delta(1-x)
  \delta^{(2)} (\vecc k_\perp) \ \mbox{\boldmath$s_{\perp}^{i}$} \ ,
\end{split}
\end{equation}
where $\lambda$ denotes the helicity and
$\mbox{\boldmath$s_{\perp}^{i}$}$
the transverse spin of the parton quark $i$.

To continue this kind of calculation beyond the tree level, we have
to expand the product of the enhanced gauge links and retain all
terms contributing up to $\mathcal{O}(g^2)$.
That is, employing the light-cone gauge
$A^+ = (A \cdot n^-) = 0$, we have to evaluate
\begin{equation}
\begin{split}
&
  [[\infty^-, \mbox{\boldmath$\infty_\perp$};
    \infty^-, \mbox{\boldmath$0_\perp$}
  ]]
\cdot
  [[\infty^-, \mbox{\boldmath$0_\perp$};
    0^-, \mbox{\boldmath$0_\perp$}
  ]]_{A^{+}=0}
=
   1 - i g \left(
               \mathcal{U}_{1} + \mathcal{U}_{2} + \mathcal{U}_{3}
         \right) \\
& ~~~~~~~~~~~~~~~~~~~~~~~~~~~~~~~~~~~~~~~~~~~~~~~~~~~~~~~~~~~~~~~~
    - g^2 \left(
                \mathcal{U}_{4} + \mathcal{U}_{5}
                + \ldots \mathcal{U}_{10}
          \right) \ ,
\end{split}
\label{eq:gauge-links-product}
\end{equation}
where the individual contributions entering this equation are
compiled in Table \ref{tab:link-terms}.\footnote{The nonlinear part
of the gluon tensor does not contribute in the considered order of
the coupling.}
In these expressions we used for the sake of convenience the
longitudinal vector $u_\mu = p^+ n^-_\mu$ which has units of mass
instead of the dimensionless light-cone vector
$n^-_\mu$, cf.\  Eq.\ (\ref{eq:rescaling}).
\begingroup
\begin{table}[t]
\caption{Individual virtual-gluon contributions appearing in
the evaluation of the product of the gauge links in Eq.\
(\ref{eq:gauge-links-product}) up to the order
$\mathcal{O}(g^2)$.
\label{tab:link-terms}}
\begin{ruledtabular}
\begin{tabular}{|c l c l|} 
Symbols & ~~~~~~~~~~~~~~~ Expressions & Figure \ref{fig:graphs} & ~~~ Value ~~~
\\ \hline \hline
$\mathcal{U}_{1}$    &  $ \int_0^\infty \! d\tau \ \vecc l \cdot \vecc {\cal A}(\vecc l \tau)$
       &  (a)
       &  $\neq 0$, \cite{CS08}                                                    \\
$\mathcal{U}_{2}$   &  $ \int_0^\infty \! d\tau \ S \cdot {\cal F}(u \tau)$
       &  (b)
       &  $\neq 0$, see text                                                       \\
$\mathcal{U}_{3}$  &  $ \int_0^\infty \! d\tau \ S \cdot {\cal F}(\vecc l \tau)$
       &  ---
       &  $0$, Eq.\ (\ref{eq:tensor-0})                                            \\
$\mathcal{U}_{4}$   &  $\int_0^\infty \! d\tau \int_0^{\tau} d \sigma
        \ (\vecc l \cdot \vecc {\cal A}(\vecc l \tau))
        \ (\vecc l \cdot \vecc {\cal A}(\vecc l \sigma))
          $
       & ---
       &  $0$, \cite{CS08}                                                         \\
$\mathcal{U}_{5}$    &  $\int_0^\infty \! d\tau \int_0^{\tau} d \sigma
        \ (\vecc l \cdot \vecc {\cal A}(\vecc l \tau))
        \ (S\cdot {\cal F}(\vecc l \sigma))
          $
       &  ---
       &  $0$, Eq.\ (\ref{eq:tensor-0})                                            \\
$\mathcal{U}_{6}$   &  $\int_0^\infty \! d\tau \int_0^{\tau} d \sigma
        \ (S\cdot {\cal F}(\vecc l \tau))\ (\vecc l
          \cdot \vecc {\cal A}(\vecc l \sigma))
          $
       &  ---
       &  $0$, Eq.\ (\ref{eq:tensor-0})                                            \\
$\mathcal{U}_{7}$  &  $ \int_0^\infty \! d\tau \int_0^{\tau} d \sigma
        \ (S\cdot {\cal F}(u \tau))\ (S \cdot {\cal F}(u \sigma))
          $
       &  (c)
       &  $0$, see text                                                            \\
$\mathcal{U}_{8}$ &  $\int_0^\infty \! d\tau \int_0^{\tau} d \sigma
       \  (S\cdot {\cal F}(\vecc l \tau))\ (S \cdot {\cal F}(\vecc l \sigma))
          $
       &  ---
       &  $0$, Eq.\ (\ref{eq:tensor-0})                                            \\
$\mathcal{U}_{9}$   &  $\int_0^\infty \! d\tau \int_0^{\infty} d \sigma
       \  (\vecc l \cdot \vecc {\cal A}(\vecc l \tau))\
          (S\cdot {\cal F} (u \sigma) )$
       &  (d)
       &  $\neq 0$, see text                                                       \\
$\mathcal{U}_{10}$    &  $\int_0^\infty \! d\tau \int_0^{\infty} d \sigma
       \  (S\cdot {\cal F}(\vecc l \tau))\ (S \cdot {\cal F}(u \sigma))
          $
       &  ---
       & $0$, Eq.\ (\ref{eq:tensor-0})                                            \\
\end{tabular}
\end{ruledtabular}
\end{table}
\endgroup

The entries in Table \ref{tab:link-terms} call for some comments and
explanations.
First, the fermion fields in the definition of the TMD PDF given by
Eq.\ (\ref{eq:TMD-PDF_fermion}) are Heisenberg field operators,
meaning that we have to use
\begin{eqnarray}
  \psi_i(\xi)
& = &
  {\rm e}^{- ig\[ \int\! d\eta \ \bar \psi \hat {\cal A} \psi \]} \
  \psi_i^{\rm free} (\xi)
  \ ,  \nonumber \\
\[ \int\! d\eta \ \bar \psi \hat {\cal A} \psi \]\
& \equiv &
  \int\! d^4\eta \ \bar \psi (\eta) \gamma_\mu
  \psi (\eta) {\cal A}^\mu (\eta) \ .
\label{eq:fermion_heis}
\end{eqnarray}
Therefore, the $\mathcal{O}(g)$ contributions
$\mathcal{U}_{1}, \mathcal{U}_{2}, \mathcal{U}_{3}$
should be contracted with the quark-gluon interaction terms
$\[ \int\! d\eta \ \bar \psi \hat {\cal A} \psi \]$,
originating from the Heisenberg fields (\ref{eq:fermion_heis}),
in order to give rise to the one-gluon exchange graphs (a) and (b),
which are of $\mathcal{O}(g^2)$.
Second, all virtual-gluon terms
$N_i\equiv \langle \mathcal{U}_{i} \rangle$ with $i=1,\ldots 10$
produce contributions of the following generic form\footnote{The
bra-ket notation used will be explained in the next section.}
\begin{equation}
\begin{split}
& \sim
  \int \! \frac{d\xi^- d^2
  \mbox{\boldmath$\xi_\perp$}}{2\pi (2\pi)^2}\,
  \exp \(- i (p^+- k^+) \xi^- \!
  +\!  i \mbox{\boldmath$k_\perp$}
  \cdot \mbox{\boldmath$\xi_\perp$} \)
  \frac{1}{2} {\rm Tr} \[ (\hat p + m)( 1 + \gamma_5 \hat s) \,
  \Gamma \ \langle N \rangle \]
  \\
& =
  \delta(p^+ - xp^+) \delta^{(2)} (\vecc k_\perp) \
  \frac{1}{2}{\rm Tr} \[ (\hat p + m)( 1 + \gamma_5 \hat s) \,
  \Gamma \ \langle N \rangle \] \ .
  \label{eq:virtual_average}
\end{split}
\end{equation}
Their Hermitean conjugated (mirror) counterparts contribute terms
of the form
\begin{equation}
\sim
  \delta(p^+ - xp^+) \delta^{(2)} (\vecc k_\perp) \
  \frac{1}{2}{\rm Tr} \[ (\hat p + m)( 1 + \gamma_5 \hat s)
  \langle N \rangle^\dag \ \Gamma \ \] \ .
\label{eq:virtual_average_conj}
\end{equation}
The Dirac structure of the quantities $\langle \mathcal{U}_{i} \rangle$
is nontrivial owing to the spin-dependent terms from the gauge
links---which we will show explicitly below.
On the other hand, the contributions of the real-gluon exchanges,
stemming from the contractions of the gauge fields belonging to
different planes in the $\xi$-space, will be considered in Sec.\
\ref{sec:real-contr}.

It is obvious that an analogous expansion has to be carried out in
Eq.\ (\ref{eq:TMD-PDF_fermion}) also for the product of the gauge
links
$
 [[\xi^-, \mbox{\boldmath$\xi_\perp$};
   \infty^-, \mbox{\boldmath$\xi_\perp$}]]^\dagger
\cdot
 [[\infty^-, \mbox{\boldmath$\xi_\perp$};
   \infty^-, \mbox{\boldmath$\infty_\perp$}]]^\dagger
$.
Let us emphasize that the various contributions of the Pauli term,
evaluated along the $n^-$-lightlike direction, do not vanish
completely when one employs the light-cone gauge---as opposed
to the standard $\sim dx_\mu A^\mu$ term.
Moreover, it was shown in \cite{JY02,BJY03,BMP03,CS07,CS08,CS09} that
the transverse gauge field in the axial gauge at light-cone infinity
is given by a total derivative, viz.,
\begin{equation}
  \vecc A^i (\infty^-, z^+, \vecc z_\perp)
=
  - \frac{1}{2} g \! \int\! \frac{dq^+ dq^-}{(2\pi)^2}
  \frac{{\rm e}^{- i q^+ \infty^- - i q^-z^+ }}{[q^+]}
  2\pi \delta (q^-) \
  \vecc \nabla^i \ \varphi (\vecc z_\perp)  \ ,
\label{eq:total-deriv}
\end{equation}
whereas the longitudinal components are equal to zero.
Thus, the field-strength tensor on the transverse segment vanishes:
\begin{equation}
  F_{a}^{\mu\nu} (\infty^-, 0^+, \vecc \xi_\perp)
=
  0 \ .
\label{eq:tensor-0}
\end{equation}
Therefore, expanding (\ref{eq:gauge-links-product}), only the terms
with {\it longitudinal} spin-dependent contributions survive, while
those with ${\cal F} (\vecc l \tau)$ (or ${\cal F} (\vecc l \sigma)$)
cancel out.
Nevertheless, we verify the vanishing of these terms by explicit
calculation in the next section.
Hence, by virtue of Eq.\ (\ref{eq:total-deriv}), expression
(\ref{eq:gauge-links-product}) reduces to
\begin{widetext}
\begin{equation}
\begin{split}
& [[\infty^-, \mbox{\boldmath$\infty_\perp$};
    \infty^-, \mbox{\boldmath$0_\perp$}]] \cdot
   [[\infty^-, \mbox{\boldmath$0_\perp$};
   0^-, \mbox{\boldmath$0_\perp$}]]_{A^{+}=0}
=
   1 - ig\! \int_0^\infty \! d\tau
   \ \vecc l \cdot \vecc {\cal A} (\vecc l \tau)
     - ig\!  \int_0^\infty \! d\tau S \! \cdot \! {\cal F} (u \tau)
\\
& ~~~~~~~~~~~~~~~~~~~~~~~~~~~~~~~~~~~~~~~~~~~~~~~~~~~~~~~~~~~~~~~~
     - g^2\! \int_0^\infty \! d\tau \int_0^{\tau} d \sigma
     \ (\vecc l \cdot \vecc {\cal A} (\vecc l \tau)) \
(\vecc l \cdot \vecc {\cal A} (\vecc l \sigma))
\\
& ~~~~~~~~~~~~~~~~~~~~~~~~~~~~~~~~~~~~~~~~~~~~~~~~~~~~~~~~~~~~~~~~
     - g^2\! \int_0^\infty \! d\tau \int_0^{\infty} d \sigma
     \ (\vecc l \cdot \vecc {\cal A} (\vecc l \tau))
       (S \cdot {\cal F} (u \sigma))
\\
& ~~~~~~~~~~~~~~~~~~~~~~~~~~~~~~~~~~~~~~~~~~~~~~~~~~~~~~~~~~~~~~~~
     + {g^2\! \int_0^\infty \! d\tau \int_0^{\tau} d \sigma
       (S \cdot {\cal F} (u \tau))
       (S \cdot {\cal F} (u \sigma))}
\\
& ~~~~~~~~~~~~~~~~~~~~~~~~~~~~~~~~~~~~~~~~~~~~~~~~~~~~~~~~~~~~~~~~
     + O(g^3)
\ .
\label{eq:gauge_links_LC}
\end{split}
\end{equation}
\end{widetext}

\section{Calculation of (virtual) gauge-field correlators}
\label{sec:virtual-corr}

We are now able to calculate the spin-dependent contributions
in Eq.\ (\ref{eq:TMD-PDF_fermion}), which we will do up to the
$g^2$-order level.
Using light-cone coordinates (also in the transverse direction), the
Pauli term reads
\begin{eqnarray}
  S \cdot {\cal F}
& \equiv &
  S_{\mu\nu} {\cal F}^{\mu \nu} \\ \nonumber
& = &
        2 S_{+-} {\cal F}^{+-}
     +  2 S_{+i} {\cal F}^{+i}
     +  2 S_{-i} {\cal F}^{-i}
     +  S_{ij} {\cal F}^{ij} \ .
\label{eq:J-F}
\end{eqnarray}
Imposing the light-cone gauge ${\cal A}^+=0$, we obtain the following
non-zero components of the field-strength tensor:
\begin{eqnarray}
  {\cal F}^{+-}
& = &
  \pd^+ {\cal A}^- \ , \ {\cal F}^{+i} = \pd^+ {\cal A}^i \ ,
\nonumber \\
  \ {\cal F}^{-i}
& = &
  \pd^- {\cal A}^i - \pd^i {\cal A}^- \ , \
  {\cal F}^{ij}
=
  \pd^i {\cal A}^j - \pd^j {\cal A}^i \ .
\label{eq:F-tensor}
\end{eqnarray}

\begin{figure*}[ht]
\centering
\includegraphics[width=0.4\textwidth,angle=90]{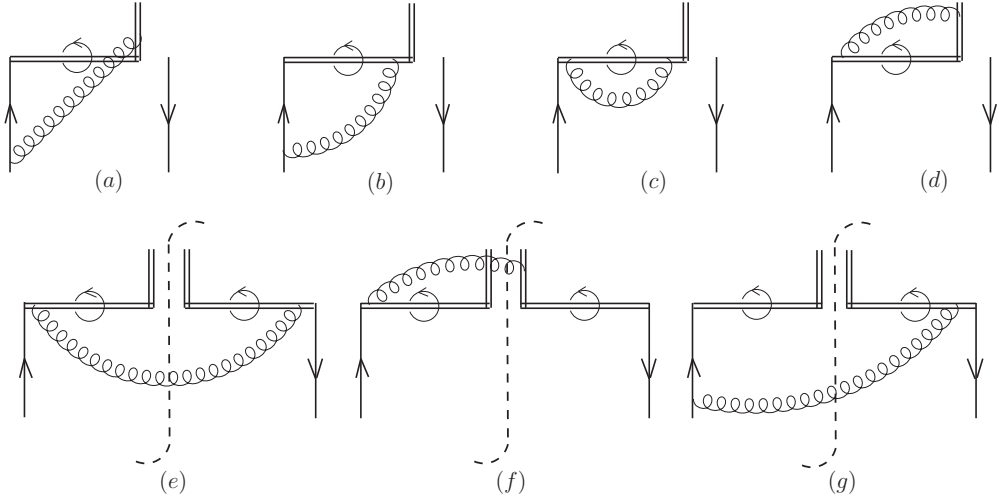}
\caption{Feynman graphs contributing to the quark-in-a-quark TMD PDF.
Double lines denote standard gauge links, while those supplemented with
a ring represent enhanced gauge links with tensor-field (Pauli)
contributions.
Fermions and gluons are shown as solid and curly lines, respectively.
Graphs (a), (b), (c), and (d) describe virtual gluon corrections;
graphs (e), (f), and (g) represent real-gluon exchanges
across the cut (vertical dashed line).
The conjugated (``mirror'') graphs are not shown for the sake
of brevity.
\protect\label{fig:graphs}}
\end{figure*}

We proceed with the explicit calculation of the \textit{virtual}
gluon exchanges in Eq.\ (\ref{eq:TMD-PDF_fermion}), relegating
the inclusion of real-gluon contributions to Sec.\
\ref{sec:real-contr}.
The reason is that only the former are UV divergent and give rise
to anomalous dimensions, while the latter contribute only UV-finite
terms.
To systematize the calculation of the various contributing
correlators, we appeal to Table \ref{tab:link-terms} in conjunction
with Fig.\ \ref{fig:graphs}.
There are two different types of contributions: those terms in
Eq.\ (\ref{eq:gauge-links-product}) which are proportional to $g^2$
stem from the evaluation of correlators between the standard gauge
links and the enhanced ones.
In Fig.\ \ref{fig:graphs} the latter are denoted by double lines
with a ring attached to them in order to indicate the Pauli
contribution which encodes spin effects.
The standard gauge links are represented by simple double lines.
The other contributions to Eq.\ (\ref{eq:gauge-links-product}),
which are proportional to $g$, i.e., the terms
$\mathcal{U}_{1}$, $\mathcal{U}_{2}$, and $\mathcal{U}_{3}$
in Table \ref{tab:link-terms}, have to be contracted with the gauge
fields generated by the Heisenberg fermion operators, cf.\
(\ref{eq:fermion_heis}), retaining again those terms which contribute
to $\mathcal{O}(g^2)$.
It is understood that each of these terms has to be averaged
over the fluctuations of the gauge field via a functional integration.
This is done with the aid of Eq.\ (\ref{eq:gluon-corr}) using in what
follows Dirac's bra-ket notation
$\langle ... \rangle$.\footnote{Strictly speaking, one should write
$\langle ... \rangle_{A}$.}

The term $\langle \mathcal{U}_{1} \rangle$---graph (a) in Fig.\
\ref{fig:graphs}---reduces in the considered order of the coupling to
what one obtains with the standard gauge links; it has been computed
in our previous work in Ref.\ \cite{CS08}.
Term $\langle \mathcal{U}_{2} \rangle$---corresponding to graph (b)
in the same figure---will be worked out below, whereas term
$\langle \mathcal{U}_{3} \rangle$
vanishes by virtue of Eq.\ (\ref{eq:tensor-0}).
For the same reason, also the contributions termed
$\langle \mathcal{U}_{5} \rangle$,
$\langle \mathcal{U}_{6} \rangle$,
$\langle \mathcal{U}_{8} \rangle$, and
$\langle \mathcal{U}_{10} \rangle$
vanish as well.
Moreover, it is proved in a few lines that
$\langle \mathcal{U}_{5} \rangle+\langle \mathcal{U}_{6} \rangle=0$.
Term $\langle \mathcal{U}_{4} \rangle$ was computed in \cite{CS08}
and was found to vanish, while the term
$\langle \mathcal{U}_{7} \rangle$, which represents the
longitudinal selfenergy contribution of the Pauli term (graph (c) in
Fig.\ \ref{fig:graphs}), will be computed further below; it amounts
again to a vanishing contribution.
Hence, the only remaining terms giving non-zero contributions are
$\langle \mathcal{U}_{2} \rangle$ and
$\langle \mathcal{U}_{9} \rangle$.
The first one stems from the interaction of the longitudinal gauge
field, produced by the fermion, with the Pauli term along the enhanced
longitudinal link---graph (b) in Fig.\ \ref{fig:graphs}---while the
second one, represented by graph (d), describes the cross talk between
the transverse gauge potential of the standard gauge link and the
longitudinal part of the Pauli term (enhanced gauge link).
Its calculation will be carried out below.
Recall that the analogous cross talk between the longitudinal parts
of the Pauli term and the standard gauge link vanishes because of
Eq.\ (\ref{eq:tensor-0}).

Having sketched the general computational framework, let us now turn a
spotlight on the calculation of the various terms, starting with
$\langle \mathcal{U}_{4} \rangle$, while the Fermion-induced terms
$\langle \mathcal{U}_{1} \rangle$, $\langle \mathcal{U}_{2} \rangle$,
and $\langle \mathcal{U}_{3} \rangle$
will be picked up in Sec.\ \ref{sec:fermion-virt}.
The term $\langle \mathcal{U}_{4} \rangle$ represents the selfenergy of
the usual transverse gauge link and vanishes in the light-cone gauge
\cite{CS08}.
This can be seen from the following equation
\begin{equation}
  \int_0^\infty \!\! d\tau \ \vecc l
  \cdot \vecc {\cal A} (\infty^-, 0^+, \vecc l \tau)
=
  \!\!\int\! \frac{dq^+}{2\pi} \frac{d^2 q_\perp}{(2\pi)^2} \ \vecc l
  \cdot \vecc \ \tilde {\cal A}(q)
  \frac{i {\rm e}^{- i q^+ \! \infty^-}}{\vecc q
  \cdot \vecc l + i0} ,
\label{eq:line-int}
\end{equation}
where we have used the Fourier transformation of the gauge field
\begin{equation}
   {\cal A}_\mu (z)
   =
   \int\! \frac{d^\omega q}{(2\pi)^\omega}
   \ {\rm e}^{-iq \cdot z} \ \tilde {\cal A}_\mu (q) \ ,
\label{eq:fourier}
\end{equation}
working in an $\omega$-dimensional momentum space
$(\omega = 4 - 2\epsilon)$.
Employing this expression in Eq.\ (\ref{eq:line-int}) and
the gluon correlator in the light-cone gauge
\begin{equation}
  \langle {\cal A}_\mu (q) {\cal A}_\nu (q') \rangle
=
  (-i) C_{\rm F} (2\pi)^4 \delta^{(4)} (q+q') \ D_{\mu\nu} (q) \ ,
\label{eq:gluon-corr}
\end{equation}
in which the regularized free gluon propagator appears
[cf.\ Eq.\ (\ref{eq:regular_pauli})],
\begin{equation}
  D_{\mu\nu} (q)
=
    \frac{2\pi \delta (q^-)}{q^2 + i0} \(g_{\mu\nu}
  - \frac{q_\mu n^-_\nu + q_\nu n^-_\mu}{[q^+]} \) \ ,
\label{eq:gluon-prop}
\end{equation}
we get
\begin{equation}
  \langle \mathcal{U}_{4} \rangle
=
  iC_{\rm F} \vecc l_\perp^2 \int\! \frac{dq^+}{2\pi}
  \int_0^\infty d\tau \int_0^\tau d\sigma \int\!
  \frac{d^2 \vecc q_\perp}{(2\pi)^2}
  \frac{{\rm e}^{- i \bit{\scriptstyle q_\perp}
  \cdot \bit{\scriptstyle l_\perp} (\tau
  - \sigma)}}{\vecc q_\perp^2 + \lambda^2}
\ .
\label{eq:transverse_SE}
\end{equation}
One notes that the gluon propagator bears a pole-prescription
dependence, codified by the symbol $[q^+]$, whereas its non-zero
parts are given by
\begin{equation}
  D_{i-}
=
  - \frac{1}{q^2 + i0} \ \frac{q^i}{[q^+]} \ , \quad
  D_{ij}
=
  i \frac{\delta_{ij}}{q^2 + i0}  \ .
\label{eq:prop_LC}
\end{equation}
Despite the pole-prescription dependence of the gluon propagator,
expression (\ref{eq:transverse_SE}) does not depend on the pole
prescription, because only the Feynman term of the gluon
propagator contributes.
Moreover, inspection of the last term in this equation reveals that
it will be canceled by its mirror contribution anyway, i.e., finally,
\begin{equation}
  \langle \mathcal{U}_{4} \rangle
=
  0 \ .
\label{eq:IV}
\end{equation}
Note that this cancelation occurs in any case: polarized or
unpolarized because there is no Dirac structure in this term.

The next two terms $\langle \mathcal{U}_{5} \rangle$ and
$\langle \mathcal{U}_{6} \rangle$,
which contain expressions of the sort
$S\cdot\mathcal{F}(\vecc l_\perp \tau)$,
can be treated in unison.
To evaluate them we make use of the derivative of the transverse
gauge field, viz.,
\begin{equation}
  {\cal A}^{i}(\vecc l_{\perp}\sigma)
=
  \int_{}^{} \frac{d^4q}{(2\pi)^4}
  {\rm e}^{- i q^{+}\infty^{-}
           + i \bit{\scriptstyle q_\perp}
  \cdot \bit{\scriptstyle l_\perp} \sigma} \!
           \vecc \ \tilde {\cal A}^{i}(q) \ .
\label{eq:fourier-A-trans}
\end{equation}
For the transverse gauge strength at light-cone infinity, one has
\begin{equation}
\begin{split}
&
  {\cal F}^{+-} (\vecc \xi_\perp)
=
  \pd^+ {\cal A}^- (\vecc \xi_\perp)
=
  0 \ , \\
&
  {\cal F}^{-i} (\vecc \xi_\perp)
=
  - \pd^i {\cal A}^- (\vecc \xi_\perp)\ , \\
&
  {\cal F}^{ij} (\vecc \xi_\perp)
=
    \pd^i {\cal A}^j (\vecc \xi_\perp)
  - \pd^j {\cal A}^i (\vecc \xi_\perp) \ ,
\label{eq:F-tensor_transverse}
\end{split}
\end{equation}
implying for the Pauli term in the transverse direction
\begin{equation}
  S \cdot \mathcal{F}(\vecc l_{\perp} \sigma)
=
  2i \int_{}^{} \frac{d^4q}{(2\pi)^4}
  {\rm e}^{- i q^{+}\infty^{-}
  + i \bit{\scriptstyle q_\perp}
  \cdot \bit{\scriptstyle l_\perp} \sigma} \! q_{\perp}^{i}
  \left[
        - S^{+i} \ \tilde{\cal A}^- (q)
        + S^{ij} \ \tilde{\cal A}^j (q)
  \right] \ .
\label{eq:J-F-trans}
\end{equation}
Then we find
\begin{eqnarray}
&&
  \langle \mathcal{U}_{5} \rangle
=
  -2 C_{\rm F} \frac{1}{T} \int_{0}^{\infty} d\tau
                           \int_{0}^{\tau} d\sigma
                           \int_{}^{} \frac{dq^+}{2\pi}
                           \int_{}^{} \frac{d^2q_\perp}{(2\pi)^2}
  {\rm e}^{-i \bit{\scriptstyle q_\perp}
  \cdot \bit{\scriptstyle l_\perp}(\tau - \sigma)}
  \frac{1}{q_{\perp}^2 + \lambda^2}
\nonumber \\
&& ~~~~~~~~~~ \times
  \left[
        -S^{+i}
        \frac{q_{\perp}^i ( \vecc q_\perp \cdot \vecc l_\perp )}{[q^+]}
        + S^{ij}q_{\perp}^i l_{\perp}^j
  \right] \ .
\label{eq:V-term}
\end{eqnarray}
An analogous calculation for the term
$\langle \mathcal{U}_{6} \rangle$ yields
\begin{equation}
  \langle \mathcal{U}_{6} \rangle = - \langle \mathcal{U}_{5} \rangle
\ ,
\label{eq:V+VI}
\end{equation}
confirming that these two contributions cancel each other.

Going forth, we can now compute the longitudinal self-energy
spin-dependent (Pauli) contribution [graph (c) in Fig.\
\ref{fig:graphs}]
\begin{equation}
  \langle \mathcal{U}_{7} \rangle
=
  \int_0^\infty \! d\tau \int_0^{\tau} d \sigma \
  (S\cdot {\cal F} (u \tau) )
  (S\cdot {\cal F} (u \sigma) )
\label{eq:Y_1_long}
\end{equation}
using
\begin{equation}
  S \cdot \mathcal{F}\left(u^{-}\tau\right)
=
  2 i \int_{}^{} \frac{d^4 q}{(2\pi)^4} {\rm e}^{-iq \cdot u \tau} q^+
  \left[
        S^{+-} \ \tilde{\cal A}^- (q) + S^{-i} \ \tilde{\cal A}^i (q)
  \right]
\label{eq:J-F-long}
\end{equation}
and employing  the regularization embodied in
Eq.\ (\ref{eq:regular_pauli}) to obtain
\begin{eqnarray}
  \langle \mathcal{U}_{7} \rangle
& = &
  - 4 i \, C_{\rm F} \, \frac{1}{T}
  \int_0^\infty \! d\tau \int_0^{\tau} d \sigma
  \int\! \frac{d^\omega q}
  {(2\pi)^\omega}\ {\rm e}^{-i q \cdot u (\tau - \sigma) }
  (q^+)^2 \
  2\pi \, \delta(q^-) \
  \frac{1}{q^2 - \lambda^2 + i0} \nonumber \\
&& \times
  \left[S^{+-} S^{+-} \,
        \frac{2 q^-}{[q^+]}
        + S^{-i} S^{+-} \, \frac{2 q^i_\perp}{[q^+]}
        - S^{-i} S^{-j} g^{ij}
  \right] \ .
\label{eq:Y_1_long_1}
\end{eqnarray}
It is easy to see that all three terms in the square bracket
give vanishing results:
The first term is zero because of $\delta (q^-)$.
The second term gives also zero due to the oddness of the
transverse integral, while the last one vanishes by virtue of
$$
 S^{-i} S^{-i}
 =
 \frac{1}{4} (\gamma^-\gamma^i)^2
 =
 0
$$
[recall that $S_{\mu\nu}=(1/4)[\gamma_{\mu}, \gamma_{\nu}]$].
Therefore, we finally get
\begin{equation}
  \langle \mathcal{U}_{7} \rangle
=
  0 \ .
\label{eq:VII-final}
\end{equation}

We consider now the term $\langle \mathcal{U}_{8} \rangle$ in more detail and
prove that it vanishes.
As we already mentioned in connection with Table \ref{tab:link-terms},
this term, which represents the self-interaction of the transverse
gauge links with the Pauli terms at light-cone infinity, vanishes by
virtue of the particular form of the transverse gauge field in the
light-cone gauge
[see Eq.\ (\ref{eq:total-deriv}) and the discussion below].
Here, we give a more detailed derivation of this result.
By definition, this term reads
\begin{equation}
  \langle \mathcal{U}_{8} \rangle
=
  \int_0^\infty \! d\tau \int_0^{\tau} d \sigma \
  (S \cdot {\cal F}(\vecc l \tau))\
  (S \cdot {\cal F}(\vecc l \sigma)) \ .
\label{eq:VIII}
\end{equation}
Therefore, we have
\begin{eqnarray}
  \langle \mathcal{U}_{8} \rangle
& = &
  - 4 i C_{\rm F} \frac{1}{T} \,
  \int_0^\infty d\tau\!  \int_0^\tau d\sigma\!
  \int\! \frac{d^\omega q}{(2\pi)^\omega} \,
  \vecc q_\perp^i \vecc q_\perp^j \
  {\rm e}^{- i \bit{\scriptstyle q_\perp}
  \cdot \bit{\scriptstyle l_\perp} (\tau - \sigma)}
  \frac{2\pi \delta(q^-) }{q^2 - \lambda^2 +i0}
\nonumber \\
&& ~ \times
  \Bigg[ S^{+i} S^{+j} \frac{2q^-}{[q^+]}
  - S^{ik} S^{+j} \frac{2q_\perp^k}{[q^+]}
  - S^{ik} S^{jl} g^{kl} \Bigg] \ .
\label{eq:J-F-J-F}
\end{eqnarray}
The term proportional to $q^-$ vanishes by virtue of the
delta-function.
The second one is equal to zero because
$S^{ik} q_\perp^i q_\perp^k = 0$.
Taking into account that the Dirac structure of the last term can
be rewritten as
\begin{equation}
  S^{ik}S^{jk} q_\perp^i q_\perp^j
=
  -\frac{1}{8} \gamma^j \gamma^i q_\perp^i q_\perp^j
=
  \frac{1}{8} \vecc q_\perp^2 \ ,
\label{eq:J-F-J-F-2}
\end{equation}
one obtains
\begin{eqnarray}
  \langle \mathcal{U}_{8} \rangle
& = &
  - 4 i C_{\rm F} \frac{1}{T} \
  \int_0^\infty d\tau \int_0^\tau d\sigma
  \int\! \frac{dq^+}{2\pi}
  \int\! \frac{d^{\omega-2} q}{(2\pi)^{\omega-2}} \,
  {\rm e}^{- q_\perp \cdot l_\perp (\tau - \sigma)}
  \ \frac{\vecc q_\perp^2 }{\vecc q_\perp^2 + \lambda^2}
\nonumber \\
& = &
  - 4 i C_{\rm F} \frac{1}{T} \ \int\! \frac{dq^+}{2\pi} \
  \int_0^\infty d\tau \int_0^\tau d\sigma
  \frac{\delta^{(\omega -2)}
                (\vecc l_\perp)}{|\tau - \sigma|^{\omega-2}}
\nonumber \\
& = &
  0
\label{eq:J-F-J-F-3}
\end{eqnarray}
in agreement with the result presented in Table \ref{tab:link-terms}.

Consider next the mixed term $\langle \mathcal{U}_{9} \rangle$, which
expresses the correlation between the longitudinal Pauli term and the
transverse gauge link [graph (d) in Fig.\ \ref{fig:graphs}], viz.,
\begin{equation}
  \langle \mathcal{U}_{9} \rangle
=
  \int_0^\infty \! d\tau \int_0^{\infty} d \sigma \
  (\vecc l \cdot {\cal A} (\vecc l \tau))
  (S\cdot {\cal F} (u \sigma) ) \ .
\end{equation}
With the help of Eqs.\ (\ref{eq:fourier}), (\ref{eq:J-F-long}), and
(\ref{eq:gluon-corr}), we obtain
\begin{equation}
\begin{split}
& \langle \mathcal{U}_{9} \rangle
=
  2 C_{\rm F} \mu^{2\epsilon} \frac{1}{T} \ \int_0^\infty \! d\tau
  \int_0^{\infty} d \sigma
  \int\! \frac{d^\omega q}{(2\pi)^\omega} \,
  {\rm e}^{iq^+\infty^- - i \bit{\scriptstyle q_\perp}
  \cdot \bit{\scriptstyle l_\perp} \sigma - i q^+ u \tau}
  2\pi \ \delta (q^-) \ q^+ \\
& ~~~~~~~~~ \times
  \[S^{+-} \vecc l^i D^{i-} (q) + S^{-j} \vecc l^i D^{ij} (q)\] \ ,
\label{eq:IX-q-plus}
\end{split}
\end{equation}
which can be recast in the form
\begin{equation}
\begin{split}
& \langle \mathcal{U}_{9} \rangle
=
  2 C_{\rm F} \mu^{2\epsilon} \frac{1}{T} \!
  \int_0^\infty \! d\tau \int_0^{\infty} d \sigma
  \int\! \frac{d^\omega q}{(2\pi)^\omega} \
  {\rm e}^{iq^+\infty^- - i \bit{\scriptstyle q_\perp}
  \cdot \bit{\scriptstyle l_\perp} \sigma - i q^+  u \tau}
  \frac{2\pi \delta (q^-)}{q^2 + i0} \\
& ~~~~~~~~~ \times
  \[S^{+-} (\vecc q \cdot \vecc l) + S^{-i} \vecc l^i q^+\]
\end{split}
\label{eq:IX-1}
\end{equation}
using Eq.\ (\ref{eq:prop_LC}).
Observe the important fact that the dependence on the pole prescription
disappeared in the above equation on account of $q^+ / [q^+] = 1$, cf.\
Eq.\ (\ref{eq:IX-q-plus}).
As a result, this equation is valid for the advanced, retarded, and
principal value prescriptions, as well as for the Mandelstam-Leibbrandt
pole prescription, though it is not obvious that it holds true in
general (see, e.g., Refs.\ \cite{CDL83,LN83,BKKN93,BA96,BHKV98}).

The $\tau$ and $\sigma$ integrations in Eq.\ (\ref{eq:IX-1})
can be performed explicitly:
\begin{eqnarray}
  \int_0^\infty \! d\tau {\rm e}^{-i q^+ u \tau}
& = &
  \frac{-i}{q^+ u - i0}, \\
  \int_0^\infty \! d\sigma
  {\rm e}^{-i \bit{\scriptstyle q_\perp}
  \cdot \bit{\scriptstyle l_\perp} \sigma}
& = &
  \frac{-i}{\vecc q \cdot \vecc l - i0} \ .
\label{eq:tau-sigma-int}
\end{eqnarray}
Making use of the following relation, which stems from the structure
of the transverse gauge field at infinity \cite{JY02,BJY03,CS08},
\begin{equation}
  \frac{{\rm e}^{iq^+ u \, \infty^-}}{q^+ u - i0}
=
  \frac{2\pi i}{u}  \ \delta (q^+) \ ,
\label{eq:transverse_delta}
\end{equation}
we get
\begin{equation}
\begin{split}
& \langle \mathcal{U}_{9} \rangle
=
   2i C_{\rm F} \mu^{2\epsilon} \frac{1}{{T} u}\, S^{+-} \!
   \int\! \frac{d^{\omega-2} q}{(2\pi)^{\omega-2}} \
   \frac{1}{\vecc q_\perp^2 +\lambda^2 - i0} \ ,
\end{split}
\label{eq:IX-2}
\end{equation}
where the ``gluon mass'' $\lambda^2$ was introduced in order to
take care of infrared singularities in the gluon propagator.
Taking into account that $T u = 1$ and performing the
$\vecc q_\perp$ integral
\begin{equation}
  \int\! \frac{d^{\omega-2} q}{(2\pi)^{\omega-2}} \
  \frac{1}{\vecc q_\perp^2 +\lambda^2 - i0}
=
  \frac{i}{4\pi} \ \(\frac{4\pi}{\lambda^2}\)^\varepsilon \
  \Gamma(\varepsilon) \ ,
\label{eq:q-perp-int}
\end{equation}
we arrive at the following final result
\begin{equation}
\begin{split}
& \langle \mathcal{U}_{9} \rangle
=
  - \frac{1}{8\pi} \
  C_{\rm F} [\gamma^+, \gamma^-] \ \Gamma (\epsilon) \
  \( 4\pi \frac{\mu^2}{\lambda^2} \)^\epsilon
\end{split}
\label{eq:J_pm}
\end{equation}
that gives rise to a UV divergence.

Its conjugated contribution, corresponding to the product
of the gauge links
\hbox{$
 [[\infty^-, \mbox{\boldmath$\xi_\perp$};
 \xi^-, \mbox{\boldmath$\xi_\perp$}]]
 [[\infty^-, \mbox{\boldmath$\infty_\perp$};
 \infty^-, \mbox{\boldmath$\xi_\perp$}]]
$},
amounts to the same expression (\ref{eq:J_pm}), i.e.,
\begin{equation}
  \langle \mathcal{U}_{9} \rangle^\dagger
= \langle \mathcal{U}_{9} \rangle
 \ .
\label{eq:J_pm_conj}
\end{equation}
But there is a crucial difference:
Now the Dirac matrix $\Gamma$ in the definition of the TMD PDF
stands on the \textit{right side} of this expression---cf.\
Eq.\ (\ref{eq:virtual_average_conj}).
Because the Dirac structure of Eq.\ (\ref{eq:J_pm}) is nontrivial,
this will lead to different results.
For instance, we get (using obvious abbreviations)
\begin{equation}
\begin{split}
(a) \ & \ \Gamma_{\rm unpol.} = \gamma^+ ~~~~~ :
    \, \Gamma_{\rm unpol.} [\gamma^+, \gamma^-]
=
    - [\gamma^+, \gamma^-] \Gamma_{\rm unpol.} \ , \\
(b)
    \ & \ \Gamma_{\rm helic.} ~ = \gamma^+ \gamma^5 ~~ :
    \, \Gamma_{\rm helic.} [\gamma^+, \gamma^-]
~ =
    - [\gamma^+, \gamma^-]\Gamma_{\rm helic.}  \ , \\
(c)
    \ & \ \Gamma_{\rm trans.} ~\! = \, i \sigma^{i+} \gamma^5 :
    \, \Gamma_{\rm trans.} [\gamma^+, \gamma^-]
=
    - [\gamma^+, \gamma^-] \Gamma_{\rm trans.}  \
\end{split}
\label{eq:gammas}
\end{equation}
From the set of these equations we conclude that after taking into
account the conjugated (mirror) contributions, all the leading-twist
two functions mutually cancel by virtue of the relation
\begin{equation}
  [\gamma^+, \gamma^-] \ \Gamma_{\rm tw-2}
=
  - \Gamma_{\rm tw-2} \ [\gamma^+, \gamma^-]
=
  2 \Gamma_{\rm tw-2} \ .
\end{equation}
This important property permits the probabilistic interpretation
of twist-two TMD PDFs, because in every term $\bar{\psi}\Gamma\psi$,
which behaves like a vector under $z$-boosts, the pole contribution
entailed by the correlation between the transverse gauge link and the
Pauli term along the longitudinal direction disappears.

Remarkably, higher-twist distribution functions (e.g., twist three),
behave differently, the reason being that they are characterized by a
different Dirac structure that remains invariant under $z$-boosts.
For example, one has for
$\Gamma_{\rm tw-3} = \gamma^i$
\begin{equation}
  [\gamma^+, \gamma^-] \ \Gamma_{\rm tw-3}
=
  \Gamma_{\rm tw-3} \ [\gamma^+, \gamma^-] \ ,
\end{equation}
so that the mutually conjugated contributions add to each other to give
the net result
\begin{equation}
  \Gamma_{\rm tw-3} \langle \mathcal{U}_{9} \rangle
+
  \langle \mathcal{U}_{9} \rangle^{\dagger} \Gamma_{\rm tw-3}
=
 -\frac{C_{\rm F}}{4\pi} [\gamma^+, \gamma^-] \Gamma (\epsilon)
  \(\! 4\pi \frac{\mu^2}{\lambda^2}\! \)^\epsilon
\label{eq:tw-3_gamma_i}
\end{equation}
making it apparent that the pole contribution in that case
is not vanishing.

The last point that has to be verified is that the term
$\langle \mathcal{U}_{10} \rangle$ in Table \ref{tab:link-terms}
vanishes.
We shall do that without assuming the special form of the gauge field
at infinity given by Eq.\ (\ref{eq:total-deriv}).
Making use of the explicit form of the gluon propagator
[cf.\ (\ref{eq:gluon-prop})], one obtains
\begin{equation}
\begin{split}
&
  \langle \mathcal{U}_{10} \rangle
=
  - 4 i C_{\rm F} \frac{1}{T} \int_0^\infty \! d\tau
                              \int_0^{\infty} d \sigma
  \int_{}^{} \frac{d^\omega q}{(2\pi)^\omega}
  {\rm e}^{ i q^+ (\infty^- - u\sigma)
           - i \bit{\scriptstyle q_\perp}
           \cdot \bit{\scriptstyle l_\perp} \tau}
  \frac{2\pi \delta (q^-) q^+ }{q^2 - \lambda^2 +i0} \\
& ~~~~~~~~~~ \times
  \[ - S^{+i}S^{+-} \frac{2 q_\perp^i q^-}{[q^+]}
     - S^{+i}S^{-j} \frac{q_\perp^i q_\perp^j}{[q^+]}
     + S^{ij}S^{+-} \frac{q_\perp^i q_\perp^j}{[q^+]}
     + S^{ij}S^{-j} q_\perp^i
  \] \ .
\label{eq:X_2}
\end{split}
\end{equation}
The first term equals zero due to the delta-function $\delta (q^-)$,
while the third one vanishes by virtue of the
antisymmetric--symmetric convolution
$S^{ij}q_\perp^i q_\perp^j = 0$.
Performing the longitudinal line integral and taking into account
Eq.\ (\ref{eq:transverse_delta}), that renders the last term
vanishing as well, we reduce the above expression to
\begin{equation}
  \langle \mathcal{U}_{10} \rangle
=
  - 4 i C_{\rm F} \frac{1}{T}  S^{+i}S^{-j}
  \int_0^{\infty} d \tau \int
  \frac{d^{\omega-2} q}{(2\pi)^{\omega-2}} \,
  {\rm e}^{i \bit{\scriptstyle q_\perp}
           \cdot \bit{\scriptstyle l_\perp} \tau}
  \frac{q_\perp^i q_\perp^j}{\vecc q_\perp^2 + \lambda^2 - i0} \ .
\label{eq:X_3}
\end{equation}
Given that
\begin{equation}
  S^{+i}S^{-j} q_\perp^i q_\perp^j
=
  \frac{1}{4} \gamma^+\gamma^- \ \vecc q_\perp^2
  \ ,
\end{equation}
we find after some standard calculations
\begin{eqnarray}
  \langle \mathcal{U}_{10} \rangle
& = &
  i C_{\rm F} \frac{1}{T} \gamma^+ \gamma^-
  \int_0^{\infty} d \tau \ \int\!
  \frac{d^{\omega-2} q}{(2\pi)^{\omega-2}} \
  \frac{\vecc q_\perp^2
  {\rm e}^{i \bit{\scriptstyle q_\perp}
  \cdot
  \bit{\scriptstyle l_\perp} \tau}}{\vecc q_\perp^2 + \lambda^2 - i0}
\nonumber \\
& = &
 0 \ .
\label{eq:X_4}
\end{eqnarray}
We thus conclude that the vanishing of the last term in
Table \ref{tab:link-terms} can be proved even without additional
constraints on the gauge field at light-cone infinity, like
Eq.\ (\ref{eq:total-deriv}).

\section{Fermion (virtual) contributions}
\label{sec:fermion-virt}

We will consider now the terms $\langle \mathcal{U}_{1} \rangle$ and
$\langle \mathcal{U}_{2} \rangle$, pertaining to graphs $(a)$ and
$(b)$ in Fig.\ \ref{fig:graphs}, and also prove that the term
$\langle \mathcal{U}_{3} \rangle$ gives zero contribution.
These are the $\mathcal{O}(g)$ terms in the expansion
(\ref{eq:gauge_links_LC}) and have to be coupled to the fermion lines
retaining their contributions up to the order $g^2$.
Term $I$ has been considered in Ref.\ \cite{CS08} and we will borrow
the result from there:
\begin{equation}
  \langle \mathcal{U}_{1} \rangle
=
  C_{\rm F} \ 2 \pi i \ C_\infty \!
  \int\! \frac{d^\omega q}{(2\pi)^\omega} \
  \frac{1}{(p-q)^2 + i0} \ \frac{\delta(q^+)}{q^2 - \lambda^2 +i0} \ ,
\label{eq:I_old}
\end{equation}
where the numerical factor $C_\infty = \{0; -1; - 1/2 \}$ corresponds
to different choices of the imposed pole-prescription in the
light-cone gluon propagator (see Refs.\ \cite{CS08, CS09}).
The UV-singularity produced by this contribution, notably,
$$
  \langle \mathcal{U}_{1} \rangle^{\rm UV}
=
  - \alpha_s \, C_{\rm F}\, \frac{1}{\varepsilon}\, i \, C_\infty
$$
just cancels the prescription-dependent term in the UV-divergent
part of the fermion selfenergy graph in the light-cone gauge,
bearing no relation to the spin-dependent part in question.

The first novel contribution, ensuing from the Pauli term,
is represented by the term $\langle \mathcal{U}_{2} \rangle$
(graph $(b)$ in Fig.\ \ref{fig:graphs})
making use of the notation we already employed in
Eq.\ (\ref{eq:gauge-links-product}).
Hence the Pauli term with the tensor gauge field in the longitudinal
direction becomes
\begin{equation}
  S\cdot {\cal F} (u \tau)
=
  2i \int\!\! \frac{d^\omega q}{(2\pi)^\omega}
  {\rm e}^{-i q \cdot u \tau}
  q^+
  \[S^{+-} \tilde {\cal A}^-(q) + S^{-i} \tilde {\cal A}^i(q)\] \ .
\label{eq:JF_1}
\end{equation}
It stems from the interaction of the quark with the spin-dependent
part of the longitudinal gauge link in the Pauli term.
Consider first its longitudinal component, which we
termed $\langle \mathcal{U}_{2}^- \rangle$, whereas for the transverse
one we will use the notation $\langle \mathcal{U}_{2}^\perp \rangle$.
Then, we have
\begin{equation}
\begin{split}
& \langle \mathcal{U}_{2}^- \rangle
=
  -  2 C_{\rm F} \
  \frac{1}{T} \int_0^{\infty}\! d\tau
  \int\! \frac{d^\omega q}{(2\pi)^\omega} \
  {\rm e}^{- i q \cdot u \tau} 2\pi \delta (q^-) q^+
  \frac{1}{(p-q)^2 + i0} \ \frac{1}{q^2 - \lambda^2 +i0} \
  \frac{1}{[q^+]} \\
& ~~~~~~~~~~ \times
    \[S^{+-} (\hat p - \hat q) \gamma^+ {2q^-}
    + S^{-i} (\hat p - \hat q) \gamma^+ {q^i_\perp} \]
  \ .
\label{eq:II-minus-step1}
\end{split}
\end{equation}
Taking into account that
$(\gamma^- \gamma^- = 0 \ , \ p^i_\perp = 0)$
\begin{eqnarray}
  S^{-i} (\hat p - \hat q) \gamma^+ {q^i_\perp}
& = &
  S^{-i} \! \[ \gamma^- \gamma^+ (p^+-q^+) q^i_\perp
  + \gamma^j \gamma^+ q^j_\perp q^i_\perp \]
\nonumber \\
& = &
  - \frac{1}{2} \gamma^- \gamma^+ \ \vecc q_\perp^2 \ ,
\label{eq:aux}
\end{eqnarray}
the pole-prescription-dependent term, containing $1/[q^+]$, cancels out
and we get
\begin{equation}
  \langle \mathcal{U}_{2}^- \rangle
=
  C_{\rm F} \gamma^- \gamma^+
  \frac{1}{T} \int_0^{\infty}\! d\tau \! \int\!
  \frac{d^{\omega - 2} \vecc q_\perp}{(2\pi)^{\omega -2}}
  \frac{dq^+}{2\pi}
  {\rm e}^{- i q \cdot u \tau}
  \frac{1}{p^2 - \vecc q_\perp^2 +i0} \ .
\label{eq:II-minus-step2}
\end{equation}
Using the fact that $u T = 1$, we finally obtain
\begin{equation}
\begin{split}
& \langle \mathcal{U}_{2}^- \rangle
=
  \frac{1}{2} C_{\rm F} \gamma^- \gamma^+ \!
  \int\! \frac{d^{\omega - 2} \vecc q_\perp}{(2\pi)^{\omega -2 }} \
  \frac{1}{p^2 - \vecc q_\perp^2 +i0}
  \ .
\label{eq:II-minus-step3}
\end{split}
\end{equation}

Turning our attention to the conjugated contribution, we find
out that the ordering of the Dirac matrices has changed:
\begin{equation}
\begin{split}
& \langle \mathcal{U}_{2}^- \rangle^\dagger
=
  -  2 C_{\rm F} \
  \frac{1}{T} \int_0^{\infty}\! d\tau
  \int\! \frac{d^\omega q}{(2\pi)^\omega} \
  {\rm e}^{- i q \cdot u \tau} 2\pi \delta (q^-) q^+
  \frac{1}{(p-q)^2 + i0} \ \frac{1}{q^2 - \lambda^2 +i0} \
  \frac{1}{[q^+]} \\
& ~~~~~~~~~~~ \times
  \[- \gamma^+ (\hat p - \hat q) S^{+-}  {2q^-}
    - \gamma^+ (\hat p - \hat q) S^{-i}  {q^i_\perp} \]
  \ .
\label{eq:II-minus-step1_conj}
\end{split}
\end{equation}
Therefore, one has
\begin{equation}
\begin{split}
&
  \langle \mathcal{U}_{2}^- \rangle^\dagger
=
  \frac{1}{2} C_{\rm F} \gamma^+ \gamma^- \!\!
  \int\! \frac{d^{\omega - 2} \vecc q_\perp}{(2\pi)^{\omega -2 }} \
  \frac{1}{p^2 - \vecc q_\perp^2  - i0}
  \ .
\label{eq:II-conj}
\end{split}
\end{equation}

Thus, combining these terms, we have for the leading-twist distribution
the following final result
\begin{eqnarray}
  \Gamma_{\rm tw-2} \langle \mathcal{U}_{2}^- \rangle
  + \langle \mathcal{U}_{2}^- \rangle^\dagger \Gamma_{\rm tw-2}
& = &
  (- i 2 \pi) \ C_{\rm F} \Gamma_{\rm tw-2}
  \int \frac{d^{2} \vecc q_\perp}{(2\pi)^{2 }}
  \delta (\vecc q_\perp^2 - p^2) \nonumber \\
& = &
  \frac{i}{2} C_{\rm F} \Gamma_{\rm tw-2} \ .
\label{eq:II-minus-final}
\end{eqnarray}
This contribution is UV finite and can be given a physical
interpretation.
Indeed, recalling Eq.\ (\ref{eq:gauge-links-product}),
we can express the above result in the form of a constant phase
\begin{equation}
  {\rm e}^{-i \delta} \approx 1 - \frac{i}{2} g^2 C_{\rm F} \ ,
\label{eq:phase}
\end{equation}
inherited to the TMD PDF by the Pauli term along the longitudinal
gauge link.
It is worth noting that this finding is valid not only for the
unpolarized case with $\Gamma_{\rm tw-2}=\gamma^+$, but also for the
polarized case ($\Gamma_{\rm tw-2}=\gamma^+\gamma^5$ or
$\Gamma_{\rm tw-2}=i\sigma^{i+}\gamma^5$).
This is because the (leading) twist-two distribution functions are
vectors under boosts along the $z$-direction and their generic Dirac
structure has the property
\begin{equation}
  \Gamma_{\rm tw-2} \gamma^-\gamma^+
=
  \gamma^+\gamma^-  \Gamma_{\rm tw-2}
=
  2 \Gamma_{\rm tw-2} \ .
\label{eq:prop}
\end{equation}
It is precisely this property that gave rise to the constant
phase $\delta$ and its cause can be traced to the correlation of
the longitudinal gauge field, concomitant to the fermions,
with the spin-dependent part of the longitudinal gauge link.
This phase could, in principle, be absorbed into the soft factor $R$
(cf.\ (\ref{eq:R-factor})).
However, this cannot be done in a universal way because the phase sign
depends on the direction of the longitudinal gauge link, i.e., on the
specific deformation of the integration contour via the $i\epsilon$
prescription.
Inverting this direction, the phase factor (\ref{eq:phase}) changes
its sign and becomes
$$
  {\rm e}^{-i \delta}
\to
  {\rm e}^{+i \delta} \ .
$$
Therefore, the phases appearing in the SIDIS and the DY process turn
out to have opposite signs:
\begin{equation}
  \delta_{\rm SIDIS}
=
  - \delta_{\rm DY} \ .
\label{eq:SIDIS-DY}
\end{equation}

These UV features do not persist for the twist-three distributions.
Indeed, their Dirac structures are invariant under $z$-boosts and
behave like scalars, i.e.,
\begin{equation}
  \gamma^-\gamma^+ \Gamma_{\rm tw-3}
\! = \!
  \Gamma_{\rm tw-3}  \gamma^-\gamma^+  , \quad\quad
  \Gamma_{\rm tw-3} \gamma^+\gamma^-
\! = \!
  \gamma^+\gamma^- \Gamma_{\rm tw-3} \ .
\label{eq:tw-3-structure}
\end{equation}
As a result, the analogous expression to (\ref{eq:II-minus-final}) now
reads
\begin{equation}
  \Gamma_{\rm tw-3} \langle \mathcal{U}_{2}^- \rangle
  + \langle \mathcal{U}_{2}^- \rangle^\dagger \Gamma_{\rm tw-3}
=
  -\frac{1}{2} C_{\rm F} \Bigg\{ \frac{1}{4\pi}
  [\gamma^+, \gamma^-]
  \(4\pi \frac{\mu^2}{p^2}\)^\epsilon \Gamma (\epsilon)
  - \frac{i}{2} \Bigg\} \Gamma_{\rm tw-3} \ .
\label{eq:tw-3-minus}
\end{equation}
This quantity is UV divergent, meaning that the Pauli spin-dependent
term will contribute to the anomalous dimension of the twist-three
TMD PDF.

We focus now on the interaction of the longitudinal spin-dependent
gauge link and the transverse part of the gauge field originating from
the fermions, namely, the term
$\langle \mathcal{U}_{2}^\perp \rangle$:
\begin{equation}
\begin{split}
&
  \langle \mathcal{U}_{2}^\perp \rangle
=
  - 2 C_{\rm F}
  \frac{1}{T} \int_0^{\infty}\! d\tau \! \int\!
  \frac{d^\omega q}{(2\pi)^\omega} \
  {\rm e}^{- i q \cdot u \tau} \ 2\pi \delta (q^-) q^+
  \frac{1}{(p-q)^2 + i0} \ \frac{1}{q^2 - \lambda^2 +i0} \
         \frac{1}{[q^+]} \\
& ~~~~~~~~~~ \times
         \[- S^{+-} (\hat p - \hat q) \gamma^j {q_\perp^j}
           + S^{-i} (\hat p - \hat q) \gamma^j q^+ g^{ij}
         \]
  \ .
\label{eq:II-perp-step2}
\end{split}
\end{equation}
Making use of the following simplifications of the terms
with Dirac matrices, i.e.,
\begin{eqnarray}
  S^{+-} (\hat p - \hat q) \gamma^j {q_\perp^j}
& = &
   S^{+-} \[ \gamma^-  (p^+ -q^+) + \gamma^+ (p^- -q^-)
   + \gamma^k q^k  \] \gamma^j q^j_\perp \nonumber \\
& \to &
  -\frac{1}{4} [\gamma^+, \gamma^-] \vecc q_\perp^2 \ ,
\end{eqnarray}
and
\begin{eqnarray}
  S^{-i} (\hat p - \hat q) \gamma^i
& = &
  S^{-i} \[ \gamma^- (p^+ -q^+)
            + \gamma^+ (p^- -q^-)
            + \gamma^k q^k  \] \gamma^i \nonumber \\
& \to &
  0  \ ,
\end{eqnarray}
and proceeding along similar lines of thought as in the previous case,
we find
\begin{equation}
\begin{split}
& \langle \mathcal{U}_{2}^\perp \rangle
=
  - \frac{1}{4} C_{\rm F}  \ [\gamma^+, \gamma^-] \!
  \int\! \frac{d^{\omega - 2} q_\perp}{(2\pi)^{\omega -2 }} \
  \frac{1}{p^2 - \vecc q_\perp^2 +i0} \
\label{eq:II-perp-step3}
\end{split}
\end{equation}
and
\begin{equation}
\begin{split}
  & \langle \mathcal{U}_{2}^\perp \rangle^\dagger
  =
   - \frac{1}{4} C_{\rm F}  \ [\gamma^+, \gamma^-] \
  \int\! \frac{d^{\omega - 2} q_\perp}{(2\pi)^{\omega -2 }} \
  \frac{1}{p^2 - \vecc q_\perp^2  - i0} \
  \ ,
\label{eq:II-perp-conj}
\end{split}
\end{equation}
so that
\begin{equation}
     \Gamma_{\rm tw-2} \langle \mathcal{U}_{2}^\perp \rangle
  +  \langle \mathcal{U}_{2}^\perp \rangle^\dagger \Gamma_{\rm tw-2}
=
     -\frac{i}{4} C_{\rm F} \ \Gamma_{\rm tw-2} \ .
\label{eq:tw-2-perp-phase}
\end{equation}
The remarks which we have made in connection with
$\langle \mathcal{U}_{2}^- \rangle$ apply equally well to
Eq.\ (\ref{eq:tw-2-perp-phase}).
The computed phase is acquired through the interaction of the Pauli
term along the longitudinal link with the transverse part of the gauge
field accompanying the fermions and has to be added to the phase
originating from the analogous interaction between the Pauli term and
the longitudinal gauge field associated to the fermions.\footnote{The
appearance of an imaginary contribution in the cusp anomalous
dimension (unrelated to the Pauli term) was already discussed by
Korchemsky and Radyushkin in \cite{KR87}.}

Thus, the full phase, ensuing from the interaction of the fermion
fields with the spin-dependent (Pauli) term in the gauge
links---crosstalk diagram $(b)$ in Fig.\ \ref{fig:graphs}---is
given according to $\langle \mathcal{U}_{2} \rangle$ by
\begin{equation}
  {\rm e}^{- i \delta}
\approx
  1 - i \, \frac{g^2}{4\pi} \, C_{\rm F} \, \pi \ ,
\label{eq:full-phase}
\end{equation}
where we have again taken into account Eq.\
(\ref{eq:gauge-links-product}).
As we have already noted, this phase flips sign when the direction
of the longitudinal link is reversed.
As regards the twist-three distribution, we obtain
\begin{equation}
  \Gamma_{\rm tw-3} \langle \mathcal{U}_{2}^\perp \rangle
  + \langle \mathcal{U}_{2}^\perp \rangle^\dagger \Gamma_{\rm tw-3}
=
  \frac{1}{4} C_{\rm F} \frac{2}{4\pi}
  [\gamma^+, \gamma^-]
  \(4\pi \frac{\mu^2}{p^2}\)^\epsilon \Gamma (\epsilon)
  \Gamma_{\rm tw-3} \ .
\label{eq:tw-3-perp}
\end{equation}
Comparison with Eq.\ (\ref{eq:tw-3-minus}) reveals that, taking their
UV divergent parts together, their total contribution disappears
leaving behind only a constant phase which, moreover, coincides with
the one found for the twist-two distributions:
$
 \delta_{\rm tw-2} = \delta_{\rm tw-3} = \alpha_{s} C_{\rm F} \pi
$.

We complete our discussion of the fermion virtual contributions
by considering the term $\langle \mathcal{U}_{3} \rangle$ in Table
\ref{tab:link-terms}, which describes the cross talk of the gauge field
surrounding the fermions with the transverse spin-dependent gauge link.
The discussion proceeds along similar lines as that of the previous
term.
Likewise, $\langle \mathcal{U}_{3} \rangle$ consists of two terms:
$\langle \mathcal{U}_{3}^- \rangle$ and
$\langle \mathcal{U}_{3}^\perp \rangle$.
Consider first the contribution pertaining to the longitudinal
gluons emanating from the quark fields:
\begin{equation}
\begin{split}
& \langle \mathcal{U}_{3}^- \rangle
=
  -  2 C_{\rm F} \
  \frac{1}{T} \int_0^{\infty}\! d\tau
  \int\! \frac{d^\omega q}{(2\pi)^\omega} \
  {\rm e}^{- i q \cdot \infty^- + i \bit{\scriptstyle q_\perp}
  \cdot \bit{\scriptstyle l_\perp} \tau}
  \frac{2\pi \delta (q^-)}{(p-q)^2 + i0} \
  \frac{1}{q^2 - \lambda^2 +i0} \
  \\
& ~~~~~~~~~~ \times
    \[- S^{+i} (\hat p - \hat q) \gamma^+
      \frac{2 q^- q_\perp^i }{[q^+]}
      + S^{ij} (\hat p - \hat q) \gamma^+
      \frac{{q^i_\perp}{q^j_\perp}}{[q^+]} \]
  \ .
\label{eq:III_long_1}
\end{split}
\end{equation}
This contribution vanishes because the first term equals zero by
virtue of the delta-function $\delta(q^-)$, while the second
one also reduces to zero due to the convolution
$S^{ij}q_\perp^i q_\perp^j$.

Continuing with the contribution from the transverse gluons
produced by the quark field, we write
\begin{equation}
\begin{split}
&
  \langle \mathcal{U}_{3}^\perp \rangle
=
  -  2 C_{\rm F} \
  \frac{1}{T} \int_0^{\infty}\! d\tau
  \int\! \frac{d^\omega q}{(2\pi)^\omega} \
  {\rm e}^{- i q \cdot \infty^- + i \bit{\scriptstyle q_\perp}
  \cdot \bit{\scriptstyle l_\perp} \tau}
  \frac{2\pi \delta (q^-)}{(p-q)^2 + i0} \
  \frac{1}{q^2 - \lambda^2 +i0} \
  \\
& ~~~~~~~ \times
  \[- S^{+i} (\hat p - \hat q) \gamma^k
  \frac{2 q_\perp^i q_\perp^k }{[q^+]}
    + S^{ij} (\hat p - \hat q) \gamma^j {q^i_\perp} \]
  \ .
\label{eq:III_1}
\end{split}
\end{equation}
After performing the following transformations of the Dirac
matrices
\begin{equation}
  S^{+i} (\hat p - \hat q) \gamma^k
\to
  S^{+i} \gamma^- \gamma^k (p^+ - q^+) - S^{+i} q_\perp^k \ ,
\end{equation}
\label{eq:dirac-trans}
and making use of Eq.\ (\ref{eq:total-deriv}), one can
recast Eq.\ (\ref{eq:III_1}) into the form
\begin{eqnarray}
  \langle \mathcal{U}_{3}^\perp \rangle
& = &
  -  4 C_{\rm F}
  \frac{1}{T} C_\infty
  \left(
        \frac{1}{i 4 \pi}
  \right)^{1-\epsilon}
  \gamma^+ \! \int_0^{\infty}\! d\tau
  \int\! \frac{d^{\omega-2} q}{(2\pi)^{\omega-2}}
  {\rm e}^{ i \bit{\scriptstyle q_\perp}
  \cdot \bit{\scriptstyle l_\perp} \tau}
  \frac{\gamma^- p^+ - \gamma^i q_\perp^i}%
  {\vecc q_\perp^2 - p^2 + i0}
\nonumber \\
& = &
    4  C_{\rm F} \
  \frac{1}{T} \frac{C_\infty}{i 4\pi} \ \gamma^+ \
  \int_0^{\infty}\! d\tau {\rm e}^{i \tau (p^2 - i0)}
  \[ \gamma^- p^+ \tau^{\varepsilon - 1/2}
  \sqrt{\frac{i\pi}{\vecc l_\perp^2}}
  + i \frac{\gamma^i l_\perp^i}{\vecc l_\perp^2} \tau^{\varepsilon -1}
  \]
  \ . ~~~
\label{eq:III_2}
\end{eqnarray}
Taking into account that $|\vecc l_\perp| \sim p^+$ and $T \sim p^+$,
one sees that both terms in the square bracket are power suppressed.
The first one, which is of ${\cal O}(|p|/p^+)$ does not diverge; the
second is of ${\cal O}(p^2/(p^+)^2)$ and is logarithmically divergent.
In any case, both terms can be left out because we are only interested
in the leading-twist contributions.
Therefore, for our analysis the correlation between the transverse
part of the Pauli term in the enhanced gauge link and the transverse
gauge field produced by the fermion can be ignored.

\section{Real-gluon contributions}
\label{sec:real-contr}

So far, we have presented the results of the calculation of the
virtual gluon graphs, which (potentially) contribute to the
UV-singularities of the TMD PDFs.
Now let us turn to the real-gluon graphs, $(e), (f), (g)$ in
Fig.\ \ref{fig:graphs}.
The formal computation of these contributions is quite similar
to that we already performed for the evaluation of the virtual graphs
$(b), (c), (d)$ in the same figure.

The main differences are: \\
(i)  The discontinuity goes now across the gluon propagator,
     so that one has to replace it with the cut one.
Then, in the light-cone gauge, we have\footnote{We omit
here the discussion of the Mandelstam-Leibbrandt pole
prescription \cite{Man82,Lei83}, making the tacit assumption
that the regularization of the $[q^+]$ pole is
$q^-$-independent.
This will allow us to avoid an additional term in the cut
propagator.}
\begin{eqnarray}
  D^{\mu\nu}
& = &
  \frac{i}{q^2 - \lambda^2 + i0} \(- g^{\mu\nu}
  + \frac{q^\mu n^{-\nu} + q^\nu n^{-\mu}}{[q^+]_{\rm PV} }\) \to
\nonumber \\
  {\rm Disc} D^{\mu\nu} (q)
& = &
    2 \pi \theta (q^+) \ \delta (q^2 -  \lambda^2)
    \( - g^{\mu\nu} + \frac{q^\mu n^{-\nu}
       + q^\nu n^{-\mu}}{[q^+]_{\rm PV} }
    \)
    \ .
\label{eq:cut_prop}
\end{eqnarray}
(ii)  The Dirac structures, abbreviated by $\Gamma$,
      stand now between Dirac matrices from the Pauli terms on
      different sides of the cut. \\
(iii) The momentum delta-functions involve, apart from
      the ``external'' momenta
      $p^+$ and $(k^+ = xp^+, \vecc k_\perp)$,
      also the ``internal'' loop momentum $q^\mu$.

The real-gluon contributions are listed in Table
\ref{tab:link-terms_real} using analogous notations to those in Table
\ref{tab:link-terms}.

\begingroup
\begin{table}[t]
\caption{Individual real-gluon contributions corresponding to
the diagrams $(e), (f), (g)$ in Fig.\ \ref{fig:graphs} and retaining
terms up to order $\mathcal{O}(g^2)$.
\label{tab:link-terms_real}}
\begin{ruledtabular}
\begin{tabular}{|c l c |} 
Symbols & ~~~~~~~~~~~~~~~ Expressions & Figure \ref{fig:graphs}
\\ \hline \hline
$\mathcal{U}_{11}$   &  $ \int_0^\infty \! d\tau \int_0^{\infty} d \sigma
        \ (S \cdot {\cal F}(u \tau))\ \Gamma \
          (S \cdot {\cal F}(u \sigma + \xi^-; \vecc \xi_\perp))
          $
       &  (e) \\
$\mathcal{U}_{12}$  &  $\int_0^\infty \! d\tau \int_0^{\infty} d \sigma
       \  (\vecc l \cdot {\cal A}(\vecc l \tau))\ \Gamma \
          (S \cdot {\cal F}(u \sigma + \xi^-; \vecc \xi_\perp))
          $
       &  (f) \\
$\mathcal{U}_{13}$ &  $ \int_0^{\infty} d \sigma
              \Gamma \ (S\cdot {\cal F}
              (u \sigma + \xi^-; \vecc \xi_\perp) )
          $
       &  (g)                      \\
\end{tabular}
\end{ruledtabular}
\end{table}
\endgroup

We start with the graph describing the interaction of two
spin-dependent gauge links, as depicted in Fig.\
\ref{fig:graphs} (e):
\begin{equation}
  \langle \mathcal{U}_{11} \rangle
=
  \int\!\frac{d\xi^- d^2 \xi_\perp}{(2\pi)^3}\
  {\rm e}^{i(p^+ - k^+) \xi^-
  - i \bit{\scriptstyle k_\perp} \bit{\scriptstyle \cdot \xi_\perp}} \!\!
  \int_0^\infty d\tau \int_0^{\infty} d \sigma
  (S\cdot {\cal F} (u \tau) ) \Gamma
  (S\cdot {\cal F} (u \sigma + \xi^-;  \vecc \xi_\perp) ) \ .
\label{eq:XI}
\end{equation}
Making use of the cut propagator (\ref{eq:cut_prop}), and
taking into account that the line integrals go along different
paths so that they have not to be ordered, one has
\begin{eqnarray}
  \langle \mathcal{U}_{11} \rangle
& = &
  - 4 \, C_{\rm F} \frac{1}{T}
  \int_0^\infty \! d\tau \int_0^{\infty} d \sigma
  \int\! \frac{d^4 q}
  {(2\pi)^4}
  {\rm e}^{-i (q^+ \cdot u  - i0 )\tau + i (q^+ u + i0)\sigma }
  \delta ((1-x)p^+) \delta^{(2)}(\vecc k_\perp - \vecc q_\perp)
\nonumber \\
&& \times
  (q^+)^2 \ 2\pi \, \delta(q^-) \ \theta (q^+)
  \delta{(q^2 - \lambda^2)}
\nonumber \\
&& \times  \Biggl[S^{+-} \ \Gamma \ S^{+-} \,
  \frac{2 q^-}{[q^+]}
  + \( S^{-i} \ \Gamma \ S^{+-}  + S^{+-}
  \Gamma \ S^{-i}\)  \, \frac{ q^i_\perp}{[q^+]}
  - S^{-i} \Gamma S^{-j} g^{ij} \Biggr]
  \ ,
\label{eq:XI_1_real}
\end{eqnarray}
noting that the dimensional regularization becomes redundant
in this case because all momentum integrals are finite due to the
delta-functions.

The $d^4q$ integration becomes trivial and hence we get
\begin{equation}
  \langle \mathcal{U}_{11} \rangle
\sim
  \[\( S^{-i} \ \Gamma \ S^{+-}  + S^{+-} \
  \Gamma \ S^{-i}\)  \, \frac{ \vecc k^i_\perp}{|1-x| p^+}
  - S^{-i} \Gamma S^{-i} \] \ .
\label{eq:XI-1}
\end{equation}
We can now make use of the following formulas
\begin{equation}
  S^{-i} \ \Gamma \ S^{+-}  + S^{+-} \ \Gamma \ S^{-i}
=
  \frac{1}{2} \gamma^i \ , \quad
  \ S^{-i} \Gamma S^{-i}
=
  \frac{1}{2} \gamma^-
\label{eq:J-aux}
\end{equation}
(assuming, for instance, $\Gamma = \gamma^+$)
and perform an averaging with the help of (\ref{eq:density}) to find
\begin{equation}
\begin{split}
&
  {\rm Tr} \[ (\hat p + m) \(1 + \gamma_5 {\hat s}\) \gamma^i \]
=
  0 \ , \\
&
  {\rm Tr} \[ (\hat p + m) \(1 + \gamma_5 {\hat s}\) \gamma^- \]
=
  2 p^- \ .
\end{split}
\label{eq:traces}
\end{equation}
From this expression one may conclude that---within the given
kinematics---the term $\langle \mathcal{U}_{11} \rangle$ is power
suppressed.

The next term represents the interaction between the longitudinal
spin-dependent gauge link and the transverse gauge link at
infinity---Fig.\ \ref{fig:graphs}, graph $(f)$.
It reads
\begin{equation}
  \langle \mathcal{U}_{12} \rangle
=
  \int_0^\infty \!\! d\tau\! \int_0^{\infty} \! \! d \sigma
  (\vecc l \cdot \vecc {\cal A} (\vecc l \tau) ) \Gamma
  (S\cdot {\cal F} (u \sigma + \xi^-; \vecc \xi_\perp) ) \ .
\label{eq:XII}
\end{equation}
The further evaluation is analogous to that in Eq.\
(\ref{eq:IX-q-plus}), giving the result
\begin{eqnarray}
  \langle \mathcal{U}_{12} \rangle
& = &
  C_{\rm F}  \frac{1}{T} \int_0^\infty d\tau
  \int_0^{\infty} d \sigma \int \frac{d^4 q}{(2\pi)^4}
  \delta((1-x)p^+ - q^+) \delta^{(2)}
  (\vecc k_\perp - \vecc q_\perp) 2\pi \ \delta (q^-) \ q^+
\nonumber \\
& \times &
  {\rm e}^{-iq^+\infty^- - i \bit{\scriptstyle q_\perp}
  \cdot \bit{\scriptstyle l_\perp} \sigma - i q^+ u \tau}
  2\pi \delta (q^2 - \lambda^2) \theta (q^+)
  \ \Gamma \,
  \[S^{+-} \vecc l\cdot \vecc q  + S^{-j} \vecc l^i q^+ \] \ .
\label{eq:XII-q-plus}
\end{eqnarray}
The last term in the square bracket disappears due to the
delta-function $\delta (q^+)$,
cf.\ Eq.\ (\ref{eq:transverse_delta}).
Thus, one gets for the sum of diagram Fig.\ \ref{fig:graphs} $(f)$ and
its mirror counterpart a vanishing contribution:
\begin{equation}
    \langle \mathcal{U}_{12} \rangle
  + \langle \mathcal{U}_{12} \rangle^\dag
\sim
  \Gamma [\gamma^+, \gamma^-] + [\gamma^+, \gamma^-] \Gamma
=
  0 \ .
\label{eq:XIII-1}
\end{equation}

\begin{figure*}[t]
\centering
\includegraphics[width=0.5\textwidth,angle=90]{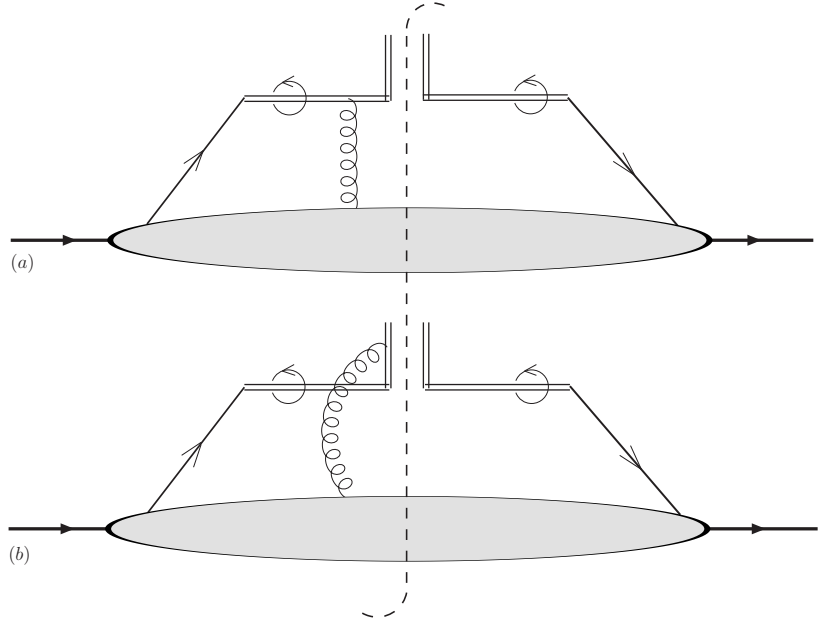}~~
\caption{Diagrams with \textbf{a} Pauli spin-dependent term, which may
contribute to the time-reversal-odd phenomena.
The intial/final-state boundary is denoted by a long vertical dashed
line.
\protect\label{fig:pauli-spectator}}
\end{figure*}

We complete our task by calculating the contribution associated
with diagram $(g)$ in Fig.\ \ref{fig:graphs}, which stems from the
interaction of the Pauli term with the fermion field.
Because it is of ${\cal O}(g)$, it has to be contracted with the
gluon field in the quark-gluon interaction term
$\[\bar \psi \hat {\cal A} \psi \]$ in order to contribute at order
$g^2$.
The combined contributions graph $(g)$ and its conjugate) of the
longitudinal part are determined by the following combinations of
Dirac matrices:
\begin{equation}
    \langle \mathcal{U}_{13}^- \rangle
  + \langle \mathcal{U}_{13}^- \rangle^\dag
\! \sim \!
  S^{-i} \Gamma (\hat p - \hat q)  \gamma^+ \!
                                 - \gamma^+ (\hat p - \hat q)
  \Gamma S^{-i} \ .
\label{eq:XIII-2}
\end{equation}
Analogously, we find for the transverse part
\begin{equation}
  \langle \mathcal{U}_{13}^\perp \rangle
  + \langle \mathcal{U}_{13}^\perp \rangle^\dag
\sim
  S^{-i} \Gamma (\hat p - \hat q) \gamma^i
        - \gamma^i (\hat p - \hat q)
         \Gamma S^{-i}
  - S^{+-} \Gamma (\hat p - \hat q) \gamma^j
  + \gamma^j (\hat p - \hat q)
    \Gamma S^{+-} \ .
\label{eq:XIII-3}
\end{equation}
After trivial manipulations with the Dirac matrices in the
equations above, we finally arrive at
\begin{equation}
    \langle \mathcal{U}_{13}^- \rangle
  + \langle \mathcal{U}_{13}^- \rangle^\dag
\sim
  xp^+ (\gamma^- \Gamma + \Gamma \gamma^-) \gamma^i
\label{eq:XIII-4}
\end{equation}
and
\begin{equation}
  \langle \mathcal{U}_{13}^\perp \rangle
  + \langle \mathcal{U}_{13}^\perp \rangle^\dag
\sim
  - xp^+ \[ (\gamma^- \Gamma + \Gamma \gamma^-) \gamma^i
           + 2 \gamma^- \Gamma \gamma^-
         \]\ .
\label{eq:XIII-5}
\end{equation}
From these results we conclude that these terms mutually cancel up
to a power-suppressed correction.

The main message from the computation of the real-gluon graphs
containing spin-dependent terms is that they do not contribute to the
TMD PDF in the leading-twist order.
All physically important effects have their roots in the contributions
of virtual-gluon exchanges.
Exactly those diagrams are responsible for time-reversal-odd effects
in more sophisticated models as we shall argue in the discussion to
follow.
\begin{figure*}[t]
\centering
\includegraphics[width=0.55\textwidth,angle=90]{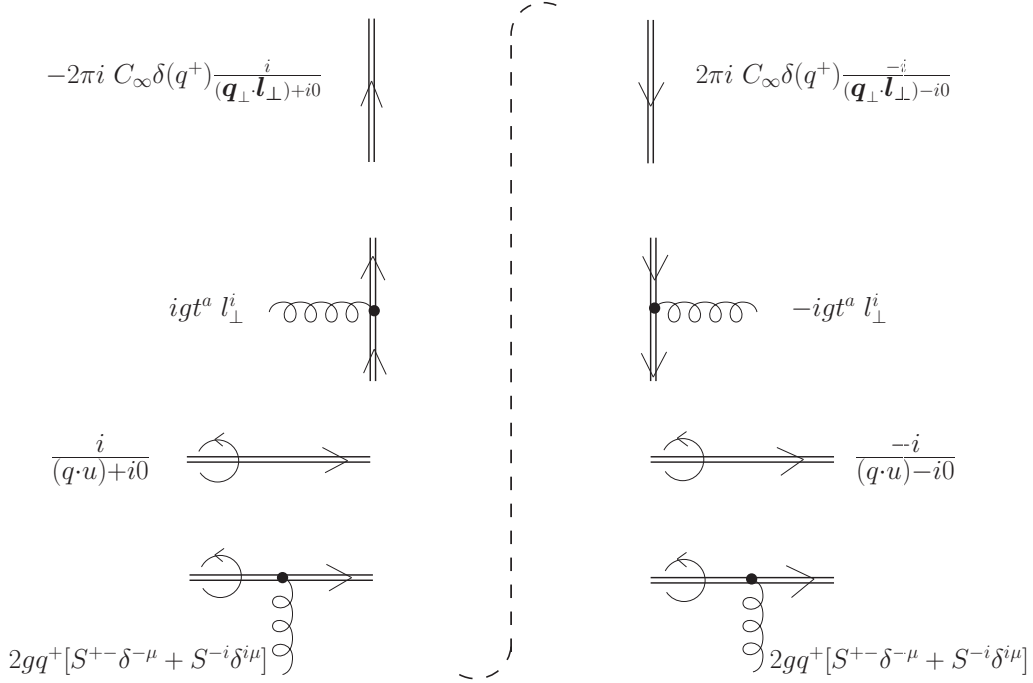}~~
\caption{The Feynman rules for the calculation of the
one-gluon-exchange graphs shown in Fig. \ref{fig:pauli-spectator} in
the light-cone gauge and using enhanced gauge links that include the
Pauli term.
Rules are given for both sides of the final-state cut (long vertical
dashed line).
Vertical double lines represent the transverse gauge links at
light-cone infinity, while the horizontal ones with arrowed rings are
the spin-dependent light-like gauge links with the Pauli terms.
\protect\label{fig:pauli-feynman}}
\end{figure*}

\section{Summary and Conclusions}
\label{sec:concl}

In this work we have presented a new gauge-invariant scheme for TMD
PDFs which takes into account in the gauge links (Wilson lines) the
Pauli term.
This term describes the communication of spin degrees of freedom with
the gauge field via the gauge-field strength and provides a rendering
geared to calculational purposes of TMD PDFs for spinning
partons.
The key features of our approach can be summarized as follows:
\begin{itemize}
  \item
  The spin-dependent Pauli term, incorporated in the TMD PDFs
  as integral part of the gauge links, does not affect their
  UV-singular behavior in leading-twist order.
  Therefore, the structure of the gauge links in the soft factor,
  introduced before in \cite{CS07,CS08} with the aim to define TMD PDFs
  in terms of gauge-invariant matrix elements with standard
  renormalization properties does not need to be changed.
  This proves the usefulness of the subtraction method, proposed in
  \cite{CH00,Hau07}, which provides a tool to deal with rapidity
  divergences that cannot be controlled by dimensional regularization.
  In fact, in our present analysis (and also in \cite{CS07,CS08}) we
  employed a soft renormalization factor in order to cure overlapping
  UV and rapidity divergences and compensate this way the associated
  one-loop cusp anomalous dimension.

  \item
  However, the Pauli term contributes to the UV-divergences of the
  imaginary parts of the cut diagrams with virtual gluon exchanges.
  Though these effects cancel in the final result of the considered
  distribution of a quark in a quark, they signalize that, within a
  more realistic context involving quark models with spectators
  \cite{GGS07,BCR08}, the spin-dependent terms may contribute to the
  interference diagrams, where these imaginary parts become crucial.

  \item
  By contrast, we found that the UV singularities of the higher-twist
  TMD PDFs (starting at twist three) are affected by the spin-dependent
  terms receiving contributions to their anomalous dimensions, which
  now become a matrix [see Eq.\ (\ref{eq:tw-3_gamma_i})].
  This is caused by an incomplete cancelation of UV divergences related
  to the fact that the $z$-boost induced by the Pauli term along the
  longitudinal link---pointing in one direction---is not counteracted
  by the conjugate contribution---pointing in the opposite direction.
  The net result is that only boosts and rotations around the
  transverse directions are left over and these give rise to a constant
  phase (see next item).
  Thus, to remedy the definition of such TMD PDFs as densities, one has
  to compensate these divergences by introducing the Pauli term also
  into the soft factor.

  \item
  An important consequence of the presence of the Pauli term in
  the gauge links is that it gives rise to a phase entanglement,
  attributable to the interaction of this spin-dependent term with the
  companion gauge field of the fermion---diagram $(b)$ in Fig.\
  \ref{fig:graphs}.\footnote{No such phase is induced by diagram $(d)$,
  the reason being that this diagram does not involve a fermion line.}
  In technical jargon, the Pauli term along the longitudinal link
  generates $z$ boosts, canceled by the conjugate link, and rotations
  along and around the transverse $x$ and $y$ directions, while the
  analogous term in the transverse gauge link produces boosts and
  rotations only along and around the transverse directions.
  The rotations of the longitudinal and the transverse link combine to
  produce a constant phase.
  It turns out that this phase correlation is the same for the leading
  twist-two and the subleading twist-three TMD PDFs, multiplying each
  of them as a whole.
  This means that absorbing this phase into the soft factor for the
  leading distribution, the corresponding phase of the subleading
  functions is also removed, even though, as we explained in the
  previous item, these latter functions may lack a density
  interpretation.
  However, the Pauli-term-induced phase is not universal because it
  depends on the direction of the longitudinal gauge link.
  Reversing the direction of the gauge link, the phase flips its sign.
  Hence, it contributes with the opposite sign to the DY process
  relative to a SIDIS situation.
  This breakdown of universality indicates that the soft
  renormalization factor $R$ does not fully decouple from the spin
  effects.
  For this to be the case, one would have to include into the
  definition of $R$ spin-dependent terms
  (cf.\ Eq.\ (\ref{eq:R-factor})) and evaluate it along a topologically
  non-trivial contour (work in progress).

\item
   To facilitate calculations with enhanced gauge links, we derive
   Feynman rules for both the left-hand side and the right-hand side
   of the final-state cut and display them in
   Fig.\ \ref{fig:pauli-feynman}.
   The spin-dependent gauge-link propagator and vertices in the
   light-cone gauge are displayed in terms of double lines with arrowed
   rings around them.
   These Feynman rules may be viewed as supplementing the set of
   Feynman rules given in \cite{CS82} for covariant gauges.
   Pay attention that the propagator of the transverse gauge link
   contains (in addition to the standard term originating from the line
   integration in momentum space) a numerical factor $C_\infty$, which
   encodes the dependence on the pole-prescription---see Eq.\
   (\ref{eq:total-deriv}).
   The cancelation of this dependence due to the soft factor is
   discussed in detail in our previous works \cite{CS07, CS08}.
\end{itemize}

Let us now close our discussion by commenting upon possible
consequences of the spin-dependent terms for models with spectators.
The time-reversal-odd TMD PDFs, like the Sivers {or the Boer-Mulders
function, which are responsible for observable single-spin
asymmetries (SSA)s, can be calculated by means of the graphs presented
in Fig.\ \ref{fig:pauli-spectator}---see for a recent analysis in
\cite{PY10}.
Such SSAs emerge as the result of the interference of the contributions
of type $(a)$ and $(b)$ with their counterparts which bear no gluon
exchanges.
For instance, within the MIT bag model, non-vanishing time-reversal-odd
TMD PDFs appear due to the interplay of the effects of the quark wave
functions in the one-gluon-interference diagrams (see, e.g., Refs.\
\cite{YU03,CDKM06,AESTYZ09,AESY10}).
Therefore, in view of our results, one may conclude that the imaginary
contributions (taken without their conjugated ``mirror'' counterparts),
which derive from the spin-dependent gauge links, can affect the
time-reversal-odd TMD PDFs even at the leading-twist level.
For instance, the Sivers function of a quark having a flavor $\alpha$
is given by
\begin{equation}
  f_{1T}^{\perp \alpha} (x, \vecc k_\perp)
\sim
    \[f_\alpha^{\gamma^+} (x, \vecc k_\perp; S_\perp)
  - f_\alpha^{\gamma^+} (x, \vecc k_\perp; - S_\perp)\]
  \ ,
\label{eq:sivers}
\end{equation}
so that it is defined by the sum of the imaginary parts of the diagrams
$(a)$ and $(b)$ in Fig.\ \ \ref{fig:pauli-spectator}.
This important finding and its phenomenological implications deserve
further exploration and verification.

\begin{acknowledgments}
We thank Anatoly Efremov for useful discussions and remarks.
This work was supported in part by the Heisenberg--Landau Program,
Grants 2009 and 2010, and the INFN.
A.~I.~K. thanks the DAAD for a research stipend at Bochum University
in the academic year 2009.
I.~O.~Ch. is grateful to Prof.\ Maxim Polyakov for the hospitality
extended to him during a visit to Bochum University, during which the
major part of this work was done, and the BMBF under Grant 06BO9012 for
financial support.
\end{acknowledgments}

\bibliographystyle{prsty-ab}

\begin{thebibliography}{[**]}

\bibitem{S76}
  D.E.~Soper,
  Phys.\ Rev.\  D {\bf 15} (1977) 1141.

\bibitem{S79}
  D.E.~Soper,
  Phys.\ Rev.\ Lett.\  {\bf 43} (1979) 1847.

\bibitem{CS81}
  J.C.~Collins, D.E.~Soper,
  Nucl.\ Phys.\ {\bf B193} (1981) 381;
  {\bf B213} (983) 545 (E).

\bibitem{CS82}
  J.C.~Collins, D.E.~Soper,
  Nucl.\ Phys.\ B {\bf 194} (1982) 445.

\bibitem{Col03}
  J.C.~Collins,
  Acta Phys.\ Pol.\ B {\bf 34} (2003) 3103.

\bibitem{TRENTO04}
  A.~Bacchetta, U.~D'Alesio, M.~Diehl, C.A.~Miller,
  Phys.\ Rev.\  D {\bf 70} (2004) 117504.

\bibitem{BR05}
  A.V.~Belitsky, A.V.~Radyushkin,
  Phys.\ Rept.\ {\bf 418} (2005) 1.

\bibitem{DM07}
  U.~D'Alesio, F.~Murgia,
  Prog.\ Part.\ Nucl.\ Phys.\  {\bf 61} (2008) 394.

\bibitem{STM10}
  P.~Schweitzer, T.~Teckentrup, A.~Metz,
  arXiv:1003.2190 [hep-ph].

\bibitem{Bacch08}
  A.~Bacchetta, D.~Boer, M.~Diehl, P.J.~Mulders,
  JHEP {\bf 0808} (2008) 023.

\bibitem{CSS89}
  J.C.~Collins, D.E.~Soper, G.~Sterman,
  Adv.\ Ser.\ Direct.\ High Energy Phys.\ {\bf 5} (1988) 1.

\bibitem{JMY04}
  X.~Ji, J.~Ma, F.~Yuan,
  Phys.\ Rev.\ D {\bf 71} (2005) 034005.

\bibitem{CRS07}
  J.C.~Collins, T.C.~Rogers, A.M.~Stasto,
  Phys.\ Rev.\ D {\bf 77} (2008) 085009.

\bibitem{CQ07}
  J.~Collins, J.W.~Qiu,
  Phys.\ Rev.\  D {\bf 75} (2007) 114014.

\bibitem{RM2010}
  T.C.~Rogers, P.J.~Mulders,
  arXiv:1001.2977 [hep-ph].

\bibitem{BMP03}
  D.~Boer, P.J.~Mulders, F.~Pijlman,
  Nucl.\ Phys.\ B  {\bf 667} (2003) 201.

\bibitem{CM04}
  J.C.~Collins, A.~Metz,
  Phys.\ Rev.\ Lett.\ {\bf 93} (2004) 252001.

\bibitem{BM07}
  C.J.~Bomhof, P.J.~Mulders,
  Nucl.\ Phys.\  B {\bf 795} (2008) 409.

\bibitem{Col08}
  J.~Collins,
  PoS {\bf LC2008} (2008) 028.

\bibitem{CS09-ER}
  I.O.~Cherednikov, N.G.~Stefanis,
  arXiv:0911.1031 [hep-ph].

\bibitem{ER80Riv}
  A.V.~Efremov, A.V.~Radyushkin,
  Riv.\ Nuovo Cim.\ {\bf 3N2} (1980) 1.

\bibitem{DGLAP}
  V.N.~Gribov, L.N.~Lipatov,
  Sov.\ J.\ Nucl.\ Phys.\ {\bf 15} (1972) 438
  [Yad.\ Fiz.\ {\bf 15} (1972) 781;
  V.N.~Gribov, L.N.~Lipatov,
  Sov.\ J.\ Nucl.\ Phys.\ {\bf 15} (1972) 675
  [Yad.\ Fiz.\ {\bf 15} (1972) 1218];
  L.N.~Lipatov,
  Sov.\ J.\ Nucl.\ Phys.\ {\bf 20} (1975) 94
  [Yad.\ Fiz.\  {\bf 20} (1974) 181];
  Y.L.~Dokshitzer,
  JETP {\bf 46} (1977) 641
  [Zh.\ Eksp.\ Teor.\ Fiz.\ {\bf 73} (1977) 1216].

\bibitem{AP77}
  G.~Altarelli, G.~Parisi,
  Nucl.\ Phys.\ B {\bf 126} (1977) 298.

\bibitem{CS08}
  I.O.~Cherednikov, N.G.~Stefanis,
  Nucl.\ Phys.\ B {\bf 802} (2008) 146.

\bibitem{CD80}
  N.S.~Craigie, H.~Dorn,
  Nucl.\ Phys.\ B {\bf 185} (1981) 204.

\bibitem{Aoy81}
  S.~Aoyama,
  Nucl.\ Phys.\ B {\bf 194} (1982) 513.

\bibitem{Ste83}
  N.G.~Stefanis,
  Nuovo Cim.\ A {\bf 83} (1984) 205.

\bibitem{JY02}
  X.~Ji, F.~Yuan,
  Phys.\ Lett.\ B {\bf 543} (2002) 66.

\bibitem{BJY03}
  A.V.~Belitsky, X.~Ji, F.~Yuan,
  Nucl.\ Phys.\ B {\bf 656} (2003) 165.

\bibitem{CS07}
  I.O.~Cherednikov, N.G.~Stefanis,
  Phys.\ Rev.\ D {\bf 77} (2008) 094001.

\bibitem{SC09}
  N.G.~Stefanis, I.O.~Cherednikov,
  Mod.\ Phys.\ Lett.\  A {\bf 24} (2009) 2913.

\bibitem{KR87}
  G.P.~Korchemsky, A.V.~Radyushkin,
  Nucl.\ Phys.\ B {\bf 283} (1987) 342.

\bibitem{CH00}
  J.C.~Collins, F.~Hautmann,
  Phys.\ Lett.\ B {\bf 472} (2000) 129.

\bibitem{Hau07}
  F.~Hautmann,
  Phys.\ Lett.\ B {\bf 655} (2007) 26.

\bibitem{CS09}
  I.O.~Cherednikov, N.G.~Stefanis,
  {Phys.\ Rev.\ D} {\bf 80} (2009) 054008.

\bibitem{Man82}
  S.~Mandelstam,
  Nucl.\ Phys.\ B {\bf 213} (1983) 149.

\bibitem{Lei83}
  G.~Leibbrandt,
  Phys.\ Rev.\ D {\bf 29} (1984) 1699.

\bibitem{Ste-unpub}
  N.G.~Stefanis, diploma thesis, Heidelberg University 1979, unpublished.

\bibitem{TM94}
  R.D.~Tangerman, P.J.~Mulders,
  Phys.\ Rev.\  D {\bf 51} (1995) 3357.

\bibitem{CDL83}
  D.M.~Capper, J.J.~Dulwich, M.J.~Litvak,
  Nucl.\ Phys.\ B {\bf 241} (1984) 463.

\bibitem{LN83}
  G.~Leibbrandt, S.L.~Nyeo,
  Phys.\ Lett.\ B {\bf 140} (1984) 417.

\bibitem{BKKN93}
  A.~Bassetto, I.A.~Korchemskaya, G.P.~Korchemsky, G.~Nardelli,
  Nucl.\ Phys.\ B {\bf 408} (1993) 62.

\bibitem{BA96}
  A.~Bassetto,
  Nucl.\ Phys.\ Proc.\ Suppl.\ {\bf 51C} (1996) 281.

\bibitem{BHKV98}
  A.~Bassetto, G.~Heinrich, Z.~Kunszt, W.~Vogelsang,
  Phys.\ Rev.\ D {\bf 58} (1998) 094020.

\bibitem{GGS07}
  L.P.~Gamberg, G.R.~Goldstein, M.~Schlegel,
  Phys.\ Rev.\  D {\bf 77} (2008) 094016.

\bibitem{BCR08}
  A.~Bacchetta, F.~Conti, M.~Radici,
  Phys.\ Rev.\  D {\bf 78} (2008) 074010.

\bibitem{PY10}
  B.~Pasquini and F.~Yuan,
  Phys.\ Rev.\  D {\bf 81} (2010) 114013.

\bibitem{YU03}
  F.~Yuan,
  Phys.\ Lett.\  B {\bf 575} (2003) 45.

\bibitem{CDKM06}
  I.O.~Cherednikov, U.~D'Alesio, N.I.~Kochelev, F.~Murgia,
  Phys.\ Lett.\  B {\bf 642} (2006) 39.

\bibitem{AESTYZ09}
  H.~Avakian, A.V.~Efremov, P.~Schweitzer, O.V.~Teryaev, F.~Yuan, P.~Zavada,
  Mod.\ Phys.\ Lett.\  A {\bf 24} (2009) 2995.

\bibitem{AESY10}
  H.~Avakian, A.V.~Efremov, P.~Schweitzer, F.~Yuan,
  arXiv:1001.5467 [hep-ph].

\end{thebibliography}

\end{document}